%% file: paper-submission.tex
\documentclass[pageno]{jpaper}

\usepackage[normalem]{ulem}

\usepackage{mathptmx} 

\newif\ifisdraft
\isdraftfalse

\input{paper-preamble}

\begin{document}
\input{paper-body}

\end{document}

%% file: paper-preamble.tex
\newcommand{\ignore}[1]{}
\usepackage{fancyhdr}
\usepackage[normalem]{ulem}
\usepackage{microtype}
\usepackage{dblfloatfix}
\usepackage[dvipsnames]{xcolor}

\usepackage{graphicx}
\usepackage{multirow,tabularx}

\usepackage{xcolor}
\usepackage{listings}
\usepackage[T1]{fontenc}
\usepackage[utf8]{inputenc}

\usepackage{hyperref}
\usepackage{pdfpages}
\usepackage{xspace}
\usepackage{subcaption}

\usepackage{enumitem}
\setlist{noitemsep} 

\usepackage{color}
\definecolor{bluekeywords}{rgb}{0.13,0.13,1}
\definecolor{greencomments}{rgb}{0,0.5,0}
\definecolor{redstrings}{rgb}{0.9,0,0}

\ifisdraft
\newcommand{\redcomment}[1]{\textcolor{red}{\textbf{#1}} }

\else
\newcommand{\redcomment}[1]{ }

\fi

\newcommand{\RA}[1]{\redcomment{RA: #1}}

\newcommand{\ei}{{\it i)}\xspace}
\newcommand{\eii}{{\it ii)}\xspace}
\newcommand{\eiii}{{\it iii)}\xspace}
\newcommand{\eiv}{{\it iv)}\xspace}

\newcommand{\isempty}[3]{%
    \if\relax\detokenize{#1}\relax
    #2%
    \else
    #3%
    \fi}

\definecolor{mGreen}{rgb}{0,0.6,0}
\definecolor{mGray}{rgb}{0.5,0.5,0.5}
\definecolor{mPurple}{rgb}{0.58,0,0.82}
\definecolor{backgroundColour}{rgb}{0.95,0.95,0.92}

\lstdefinestyle{CStyle}{
  backgroundcolor=\color{backgroundColour},   
  commentstyle=\color{mGreen},
  keywordstyle=\color{magenta},
  numberstyle=\tiny\color{mGray},
  stringstyle=\color{mPurple},
  basicstyle=\scriptsize\ttfamily,
  breakatwhitespace=false,         
  breaklines=true,                 
  captionpos=b,                    
  keepspaces=true,                 
  showspaces=false,                
  showstringspaces=false,
  showtabs=false,                  
  tabsize=2,
  language=C,
  escapeinside={(*@}{@*)},
  morekeywords={uint8_t, uint16_t, uint32_t, uint64_t, size_t},
}

\lstdefinestyle{CStyleInline}{
  backgroundcolor=\color{backgroundColour},   
  commentstyle=\color{mGreen},
  keywordstyle=\color{magenta},
  numberstyle=\tiny\color{mGray},
  stringstyle=\color{mPurple},
  basicstyle=\footnotesize\ttfamily,
  breakatwhitespace=false,         
  breaklines=true,                 
  captionpos=b,                    
  keepspaces=true,                 
  showspaces=false,                
  showstringspaces=false,
  showtabs=false,                  
  tabsize=2,
  language=C,
  escapeinside={(*@}{@*)},
  morekeywords={uint8_t, uint16_t, uint32_t, uint64_t, size_t},
}

\newcommand{\codeinline}[1]{\lstinline[style=CStyleInline]{#1}}

%% file: paper-body.tex

\newcommand{\sys}{\emph{Mitosis}\xspace}

\title{Mitosis: Transparently Self-Replicating Page-Tables for Large-Memory Machines\vspace{-0.6cm}}
\author{{Reto Achermann\textsuperscript{1,2}\quad Ashish Panwar\textsuperscript{1,3}\quad Abhishek Bhattacharjee\textsuperscript{4}\quad
        Timothy Roscoe\textsuperscript{2}\quad Jayneel Gandhi\textsuperscript{1}}\\
        \normalsize{\textsuperscript{1}VMware Research\quad \quad
        \textsuperscript{2}ETH Zurich\quad \quad \textsuperscript{3}IISc Bangalore\quad \quad \textsuperscript{4}Yale University}\\
	\normalsize{\emph{reto.achermann@inf.ethz.ch, ashishpanwar@iisc.ac.in, abhishek@cs.yale.edu, troscoe@inf.ethz.ch, gandhij@vmware.com}}\vspace{-0.4cm}}
\date{}
\maketitle

\thispagestyle{empty}


\input{00-abstract.tex}
\input{11-introduction}

\input{20-background}
\input{30-analysis}

\input{40-design}

\input{51-discussion}
\input{60-evaluation}

\input{80-conclusion}

\RA{THIS SHOULD BE AT 11 PAGES!!!}

%
%
\bibliographystyle{IEEEtranS}
\bibliography{bibliography}

%% file: 00-abstract.tex
\begin{abstract}

Multi-socket machines with 1-100 TBs of physical memory are becoming 
prevalent. Applications running on multi-socket
machines suffer non-uniform bandwidth and latency when accessing physical
memory. Decades of research have focused on data 
allocation and placement policies in NUMA settings, but there
have been no studies on the question of how to place page-tables
amongst sockets. We make the case for 
explicit page-table allocation policies and show that page-table placement is 
becoming crucial to overall performance.

We propose \sys to mitigate NUMA effects on page-table walks by
transparently replicating and migrating page-tables across
sockets without application changes. This reduces the frequency of accesses 
to remote NUMA nodes when performing page-table walks.
\sys uses two components: (i) a mechanism to enable efficient page-table
replication and migration; and (ii) policies for processes to
efficiently manage and control page-table replication and
migration. 

We implement \sys in Linux and evaluate its benefits 
on real hardware. \sys improves performance for large-scale multi-socket 
workloads by up to 1.34x by replicating page-tables across sockets. 
Moreover, it improves performance by up to 3.24x in cases when the OS migrates 
a process across sockets by enabling cross-socket page-table migration.
\end{abstract}

%% file: 11-introduction.tex
\section{Introduction}
\label{sec:introduction}

In this paper, we investigate the performance issues in large NUMA
systems caused by the sub-optimal placement not of program data, but
of page-tables, and show how to mitigate them by replicating
and migrating page-tables across sockets.

The importance of good data placement across sockets for performance on NUMA 
machines is well-known 
~\cite{Calciu:2017:BCD,Dashti:2013,Gaud:2014:LPM,Kaestle:2015:SSA}. However, the 
increase in main memory size is outpacing the growth of TLB capacity. Thus, 
TLB coverage (i.e. the size of memory that TLBs map) is stagnating and 
is causing more TLB misses 
~\cite{Basu:2013:dir_seg,Karakostas:2015:rmm,Pham:2014,Pham:2012}.
Unfortunately, the performance penalty due to a TLB miss is significant (up 
to 4 memory accesses on x86-64). Moreover, this penalty will grow to 5 memory 
accesses with Intel’s new 5-level page- tables ~\cite{intel:5level}. 

\begin{figure}[t]
  \centering
  \includegraphics[width=\columnwidth]{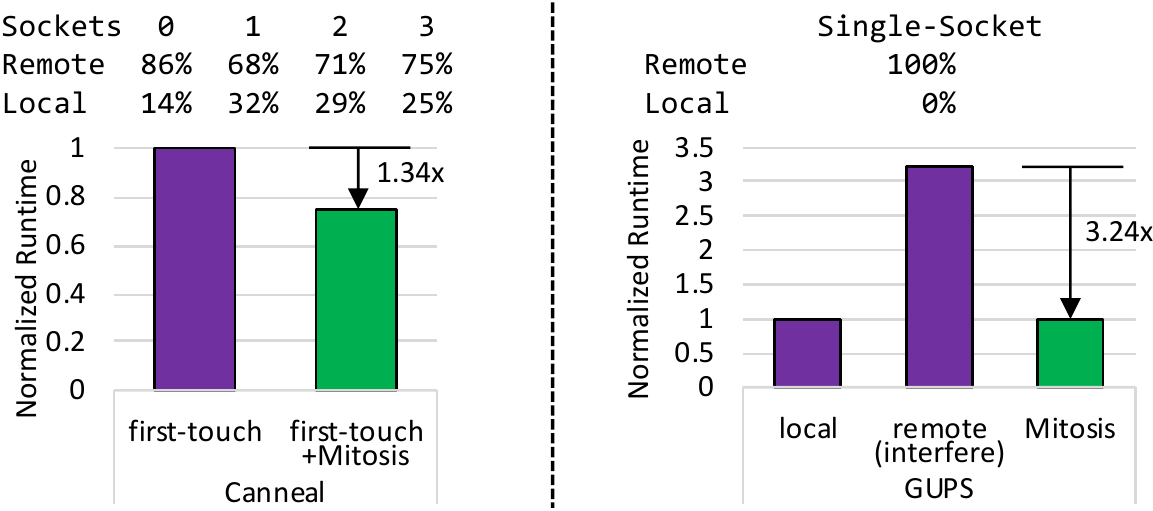}
  \caption{\emph{Top Table:} Percentage of local and remote leaf PTEs as observed from each socket on a TLB miss and \emph{Bottom Graph:} Normalized runtime, for two workloads showing multi-socket (left) and workload migration (right) scenarios with their respective improvement using \sys.} 
\label{fig:intro:pagetablewalks}
\end{figure}

Our \underline{first} contribution in this paper (\autoref{sec:overheads}) is to
show by experimental measurements on a real system that \emph{page-table} placement in
large-memory NUMA machines poses performance challenges: 
a page-table walk may require multiple remote DRAM
accesses on a TLB miss and such misses are increasingly frequent.
We show this effect due to page-table placement on a large-memory machine in two scenarios.
The first is a \emph{multi-socket scenario (\autoref{sec:scenario:multi})}, where large-scale multithreaded workloads execute across all sockets.
In this case, the page-table is distributed across sockets by the OS as it sees fit.
Such page placement results in multiple remote page-table accesses, degrading performance. 
We show the percentage of remote/local page-table entries (PTEs) on a TLB miss as observed from each socket in the top left table 
of~\autoref{fig:intro:pagetablewalks} for one workload (Canneal) from the multi-socket scenario. 
We observe that some sockets experience longer TLB misses since up to 86\% of leaf PTEs are located remotely.
Large-memory workloads like key-value stores and databases that stress TLB capacity are particularly susceptible to this behavior.

Our second analysis configuration focuses on a \emph{workload migration scenario (\autoref{sec:scenario:single}}), where the OS decides to
migrate a workload from one socket to another. Such behavior arises for many reasons:
the need to load balance, consolidate, improve cache behavior, or save power/energy~\cite{vmware:numa, Das:Albatross, Rangan:Thread}. 
A key question with migration is what happens to the data that the workload accesses.
Existing NUMA policies in commodity OSes migrates data pages to the target socket where the workload has been migrated.
Unfortunately, page-table migration is not supported~\cite{ashish:2018}, making future TLB misses expensive. 
Such misplacement of page-tables leads to performance degradation for the workload since 100\% of TLB misses require remote memory access
as shown in top right table of ~\autoref{fig:intro:pagetablewalks} for one workload (GUPS) from workload migration scenario. 
Workload migration is common in environments where virtual machines or containers are consolidated on large systems~\cite{vmware:numa}.
Ours is the first study to show this problem of sub-optimal page-table placement on NUMA machine using these two commonly occurring scenarios.

Our \underline{second} contribution (\autoref{sec:design:concept}) is a technique, \sys,
which replicates and migrates page-tables to reduce this effect. \sys
works entirely within the OS and requires no change to application
binaries. The design consists of a mechanism to enable efficient page-table replication 
and migration (\autoref{sec:design:mechanism}), and associated policies for processes to
effectively manage page-table replication and migration (\autoref{sec:design:policy}).
\sys builds on widely-used OS mechanisms like page-faults and system calls
and is hence applicable to most commodity OSes.

Our \underline{third} contribution (\autoref{sec:design:mechanism},~\ref{sec:design:policy}) is an
implementation of \sys for an x86-64 Linux kernel. 
Instead of substantially re-writing the memory subsystem, 
we extend the Linux PV-Ops~\cite{pvops} interface to page-tables
and provide policy extensions to Linux's standard user-level NUMA
library, allowing users to control migration and replication of page-tables, 
and selectively enable it on a per-process basis.
When a process is scheduled to run on a core, we load the core's
page-table pointer 
with the physical address of the local page-table replica for the socket. When
the OS modifies the page-table, the updates are propagated
to all replicas efficiently and that page-table reads
return consistent values based on all replicas. 

An important feature of \sys is that it requires
no changes to applications or hardware, and is easy to use on a per-application basis. 
For this reason, \sys is readily deployable and complementary
to emerging hardware techniques to reduce address translation
overheads like segmentation~\cite{Basu:2013:dir_seg,Karakostas:2014}, PTE coalescing~\cite{Pham:2014,Pham:2012} and user-managed virtual memory~\cite{Alam:2017}. 
We will release our implementation of \sys to enable 
future research on page-table placement and plan to upstream our changes to Linux.

Our \underline{final} contribution (\autoref{sec:evaluation}) is a 
performance evaluation of \sys on real hardware.  We show the effects of page-table
replication and migration on a large-memory machine in the same two scenarios used before to analyze page-table placement. 
In the first, \emph{multi-socket scenario}, we had observed that page-table placement results in multiple remote memory accesses,
degrading performance for many workloads. 
The graph on the bottom left of~\autoref{fig:intro:pagetablewalks} shows the performance of
a commonly used ``first-touch'' allocation policy which allocates data pages local to the socket that touches
the data first. This policy is not ideal as it cannot allocate page-tables locally for all sockets. \sys replicates
page-tables across sockets to improve performance by up to 1.34x in this scenario.
These gains come at a mere cost of 0.6\% memory overhead compared to the exorbitant memory cost of data replication.

In the second, \emph{workload migration scenario}, we had observed that page-table migration is not supported, 
which makes TLB misses expensive for workloads after their migration across sockets.
The graph on the bottom right in~\autoref{fig:intro:pagetablewalks} quantifies the worst-case performance impact of
misplacing page-tables on memory that is remote with respect to the application socket (see remote (interfere) bar).
The local bar shows the ideal execution time with locally allocated page-tables. \sys improves this situation
by enabling cross-socket page-table migration, and boosts performance by up to 3.24x.

%% file: 20-background.tex
\section{Background}
\label{sec:background}

\subsection{Virtual Memory}

Translation Lookaside Buffers (TLBs) enable fast 
address translation and are key to the performance of a virtual memory based system.
Unfortunately, TLBs only cover a tiny fraction of physical memory available on modern systems
while workloads consume all memory for storing their large datasets. 
Hence, memory-intensive workloads incur frequent costly TLB misses requiring page-table lookup by hardware.

Research has shown that TLB miss processing is prohibitively 
expensive~\cite{Basu:2013:dir_seg,Bhattacharjee:2013, Bhattacharjee:2017, 
Bhattacharjee:2011,Gandhi:2016:micro, Lustig:2013} as walking page-tables (e.g., 4-level radix tree on x86-64)
requires multiple memory accesses. Even worse, virtualized systems need 
two-levels of page-table lookups
which can result in much higher TLB miss processing overheads (24 memory accesses instead of four on x86-64). Consequently, address translation
overheads of 10-40\% are not unusual~\cite{Basu:2013:dir_seg, Bhattacharjee:2013,Bhattacharjee:2017, 
Gandhi:2014:emv,Gandhi:2016:agile,Karakostas:2015:rmm}, and will worsen with emerging 5-level page-tables \cite{intel:5level}.

In response, many research proposals improve address translation by reducing the frequency of
TLB misses and/or accelerating page-table walks. Use of large pages to increase TLB-coverage~\cite{Du:2015,Fang:2001,Ganapathy:1998,Navarro:2002,Papadopoulou:2015,
Seznec:2004,Swanson:1998,Talluri:1994} and
additional MMU structures to cache multiple levels of the page-tables~\cite{Barr:2010,Bhattacharjee:2013,Bhattacharjee:2011} are some of the techniques
widely adopted in commercial systems. In addition, researchers have also proposed TLB-speculation~\cite{Barr:2011,Pham:2015},
prefetching translations~\cite{Kandiraju:2002,Lustig:2013,Saulsbury:2000}, eliminating or devirtualizing virtual memory~\cite{Haria:2018}, or exposing
virtual memory system to applications to make the case for application-specific address translation~\cite{Alam:2017}.

We observe that prior works studied address translation on single-socket systems. However,
page-tables are often placed across remote and local memories in large-memory
systems. Given the sensitivity of large page placement on
such systems~\cite{Gaud:2014:LPM}, we were intrigued by the question of how page-table
placement affects overall performance. In this paper, we present compelling
evidence to show that optimizing page-table placement is as crucial as optimizing data placement.

\subsection{NUMA Architectures}

Multi-socket architectures, where CPUs are connected via a cache-coherent interconnect,
offer scalable memory bandwidth even at high capacity and are frequently used in
modern data centers and cloud deployments.
Looking forward, this trend will only increase; large-memory (1-100 TBs) machines are
integrating even more devices with different performance characteristics like Intel's Optane
memory~\cite{intel:3dxpoint}. Furthermore,  emerging architectures using chiplets and multi-chip modules
~\cite{amd:nextgen,Demir:2014,intel:amd,Iyer:2016,Kannan:2015,mochi,tsmc:packaging,
 Yin:2018:chiplet} will drive the multi-socket and NUMA paradigm: 
accessing memory attached to the local socket will have higher bandwidth and
lower latency than accessing memory attached to a remote socket. 
Note that accessing remote memory can incur 2-4x higher latency than accessing
local memory~\cite{latency}. Given the non-uniformity of access latency and bandwidth, optimizing
data placement in NUMA systems has been an active area of research.

\subsection{Data Placement in NUMA machines}

Modern OSes provide generic support for optimizing data placement on NUMA systems
through various allocation and migration polices. For example, Linux provides first-touch vs.
interleaved allocation to control the initial placement of data, and additionally employs
AutoNUMA to migrate pages across sockets in order to place data closer to the threads
accessing it. To further optimize data placement, Carrefour~\cite{Dashti:2013} 
proposed data-page replication
along with migration. In addition, data replication has also been proposed at data structure
level~\cite{Calciu:2017:BCD} and via NUMA-aware memory allocators~\cite{Kaestle:2015:SSA} to
further reduce the frequency of remote memory accesses. In contrast, our work focuses on
page-table pages, not data pages.

Some prior research has proposed replicated data structures for address spaces.
RadixVM \cite{Clements:2013} manages the process' address
space using replicated radix trees to improve the scalability of virtual
memory operations 
in the research-grade xv6 OS~\cite{xv6}. However, \emph{RadixVM} does not
replicate page-tables.
Similarly, Corey~\cite{Boyd-Wickizer:2008} divides the address space into shared
and private per-core regions where these explicitly shared regions
share the page-table. In contrast, we use replication to manage NUMA effects of page-table
walks in an industry-grade OS.

\noindent
{\bf Techniques for data vs. page-table pages:} One may expect prior
migration and replication  techniques to extend readily to page-tables. 
In reality, subtle distinctions between data and page-table pages merit some discussion. 
First, data pages are replicated by simple \emph{bytewise} copying of data, 
without any special reasoning of the contents of the pages. 
Page-table pages, however, require more care and cannot rely simply on 
bytewise copying -- to semantically replicate virtual-to-physical mappings, 
upper 
page-table levels must hold pointers (physical addresses) to their replicated, 
lower level page-tables -- which differ from replica to replica except at the leaf level.
Moreover, data replication has high memory overheads and maintaining
consistency across replicated pages (especially for write-intensive pages) can
outweigh the benefits of replication. While data replication has its values, we show that
page-table replication is equally important -- it incurs negligible memory overhead, can be
implemented efficiently and delivers substantial performance improvement.

%% file: 30-analysis.tex
\section{Page-Table Placement Analysis}
\label{sec:overheads}

In this section, we first present an
analysis of page-table distributions when running memory-intensive 
workloads on a large-memory machine (\emph{multi-socket scenario}~\autoref{sec:scenario:multi}) and then quantify the impact of
NUMA effects on page-table walks (\emph{workload migration scenario}~\autoref{sec:scenario:single}). Our experimental platform is
a 4-socket Intel Xeon E7-4850v3 with 512 GB physical memory (more detailed machine configuration in~\autoref{sec:evaluation}).

\subsection{Multi-Socket Scenario}
\label{sec:scenario:multi}

\begin{figure}[t]
\begin{center}
\includegraphics[width=0.75\columnwidth]{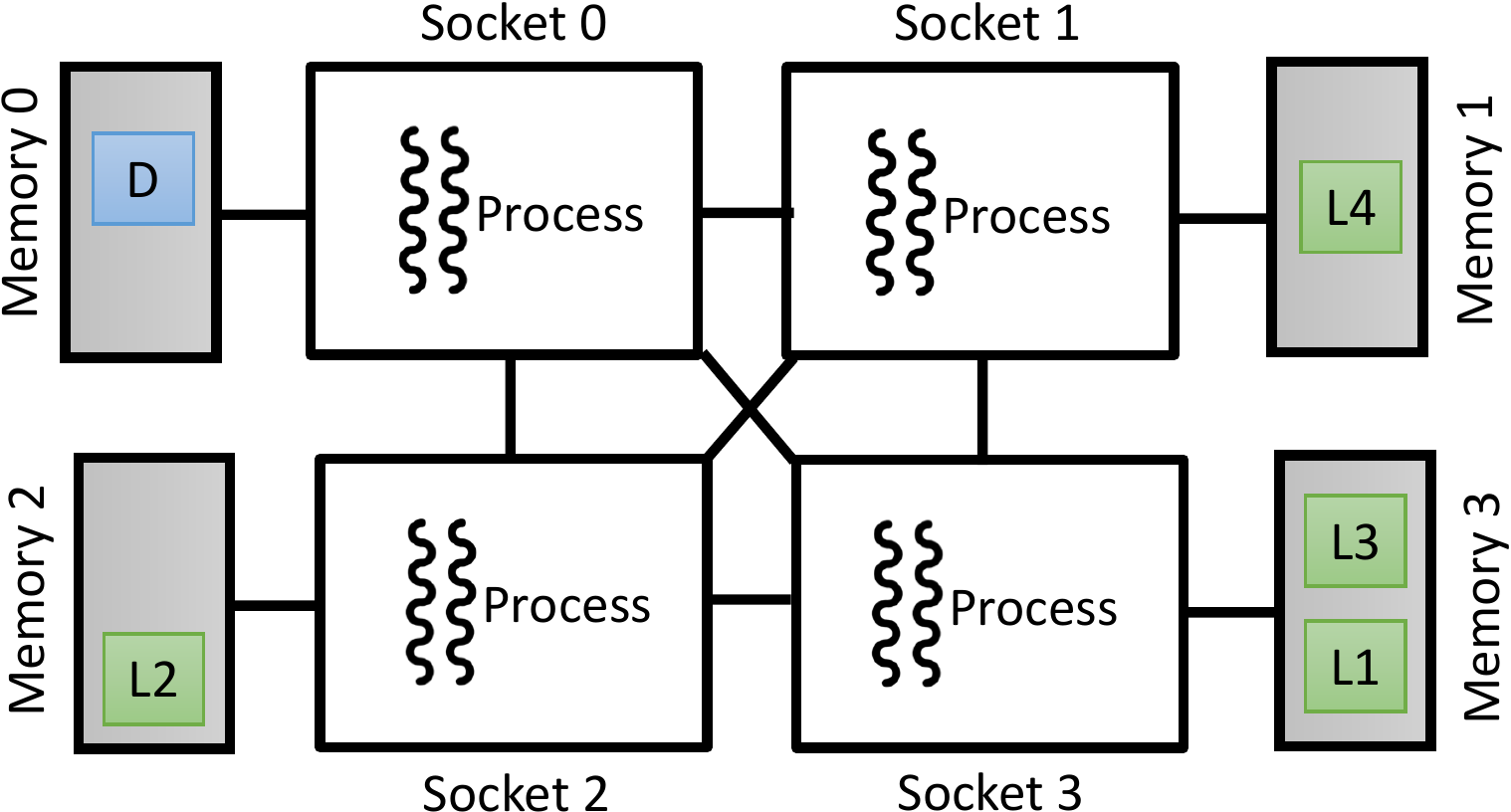}
\end{center}
	\vspace{-0.3cm}
	\caption{An illustration of current page-table and data placement for a multi-socket workload using 4-socket system.}
	\label{fig:migration:example}
\end{figure}

We focus on page-table distributions where workloads use almost all 
resources in a multi-socket system. Consider the example in~\autoref{fig:migration:example}.
If a core in socket 0 has a TLB miss for data "D", which is local to the socket, it has to perform up to 4 remote accesses
to resolve the TLB miss to ultimately discover that data was actually local to its socket.
Even though MMU caches~\cite{Barr:2010} help reduce some of the accesses, at least leaf-level
PTEs have to be accessed. Since big-data workloads have large page-tables that are absent from
the caches, system memory accesses are often unavoidable~\cite{Bhattacharjee:2017}.

\noindent
{\bf Methodology.} We are interested in the distribution of pages for
each level in the page-table; i.e., which sockets page-tables are
allocated on. We write a kernel module that
walks the page-table of a process and dumps the PTEs
including the value of the page-table root register (CR3) to a
file. The kernel module is then invoked every 30 seconds while a
multi-socket workload (e.g., Memcached) ran, producing a
stream of page-table snapshots over time. We use 30 second time intervals
as page-table allocation occurs relatively infrequently and 
smaller time interval does not change results significantly. 
We use first-touch or interleaved data allocation policy while 
enabling/disabling AutoNUMA~\cite{autonuma}
data page migration with different page sizes for multi-socket workloads in~\autoref{tab:workload}.

\begin{table}[b!]
  \begin{center}
    \begin{scriptsize}
            \begin{tabular}{m{1.2cm}m{5cm}m{0.8cm}m{0.8cm}} \\ \hline
                    {\bf Workload} & {\bf Description} & {\bf MS} & {\bf WM}\\ \hline
                    Memcached & a commercial distributed in-memory object caching system~\cite{memcached} & 350GB & -- \\ \hline
                    Graph500 & a benchmark for generation, compression and search of large graphs~\cite{graph500} & 420GB & -- \\ \hline
                    HashJoin & a benchmark for hash-table probing used in database applications and other large applications & 480GB & 17GB  \\ \hline
                    Canneal & a benchmark for simulated cache-aware annealing to optimize routing cost of a chip design~\cite{parsec} & 382GB & 32GB\\ \hline
                    XSBench & a key computational kernel of the Monte Carlo neutronics application~\cite{xsbench} & 440GB & 85GB\\ \hline
                    BTree & a benchmarks for index lookups used in database and other large applications & 145GB & 35GB\\ \hline
                    LibLinear & a linear classifier for data with millions of instances and features~\cite{svm} & -- & 67GB\\ \hline
                    PageRank &  a benchmark for page rank used to rank pages in search engines~\cite{gapbs} & -- & 69GB \\ \hline
                    GUPS & a HPC Challenge benchmark to measure the rate of integer random updates of memory~\cite{gups} & -- & 64GB\\ \hline
                    Redis & a commercial in-memory key-value store~\cite{redis} & -- & 75GB\\ \hline
      \end{tabular}
    \end{scriptsize}
  \end{center}
	\vspace{-0.3cm}
        \caption{Workloads used for analysis in multi-socket (MS) and workload migration (WM) scenarios.}
        \label{tab:workload}
\end{table}

\begin{figure*}[ht!]
    \begin{scriptsize}
        \begin{center}
      \begin{verbatim}
  Level |          Socket 0           |         Socket 1            |           Socket 2          |           Socket 3
    L4  |   0 [  0   0   0   0] ( 0%) |   1 [  8   3   0   1] (75%) |   0 [  0   0   0   0] ( 0%) |   0 [  0   0   0   0] ( 0%)
    L3  |   1 [ 56  66  40  37] (72%) |   3 [ 33  43  26  26] (66%) |   0 [  0   0   0   0] ( 0%) |   0 [  0   0   0   0] ( 0%)
    L2  |  89 [11k 11k 11k 11k] (75%) | 109 [13k 13k 13k 13k] (75%) |  66 [ 8k  8k  8k  8k] (75%) |  63 [ 7k  7k  7k  8k] (75%)
    L1  | 40k [ 6M  4M  4M  4M] (67%) | 40k [ 4M  6M  4M  4M] (67%) | 40k [ 4M  4M  6M  4M] (67%) | 40k [ 4M  4M  4M  6M] (67%)
\end{verbatim}
        \end{center}
    \end{scriptsize}
	\vspace{-0.3cm}
	\caption{Analysis of page-table pointers from a page-table dump for a multi-socket workload: Memcached.}
	\label{fig:pgtable:ptrs}
\end{figure*}

\noindent
{\bf Analysis.} We analyze the distribution of page-tables for each snapshot
in time. For each page-table level, we summarize the number of
per-socket physical pages and the number of valid PTEs pointing to
page-table pages (or data frames) residing on a local or remote socket. 
From these snapshots, we collect a distribution of leaf
PTEs and which sockets they are located on. We focus on leaf PTEs as there are orders of magnitude more of them
than non-leaf PTEs and because they generally determine 
address translation performance (upper-level PTEs can be cached in MMU caches \cite{Bhattacharjee:2017}).
These distributions indicate how many local and remote sockets a page-table
walk may visit before resolving a TLB miss.

\begin{figure}[t]
\begin{center}
\includegraphics[width=1\columnwidth]{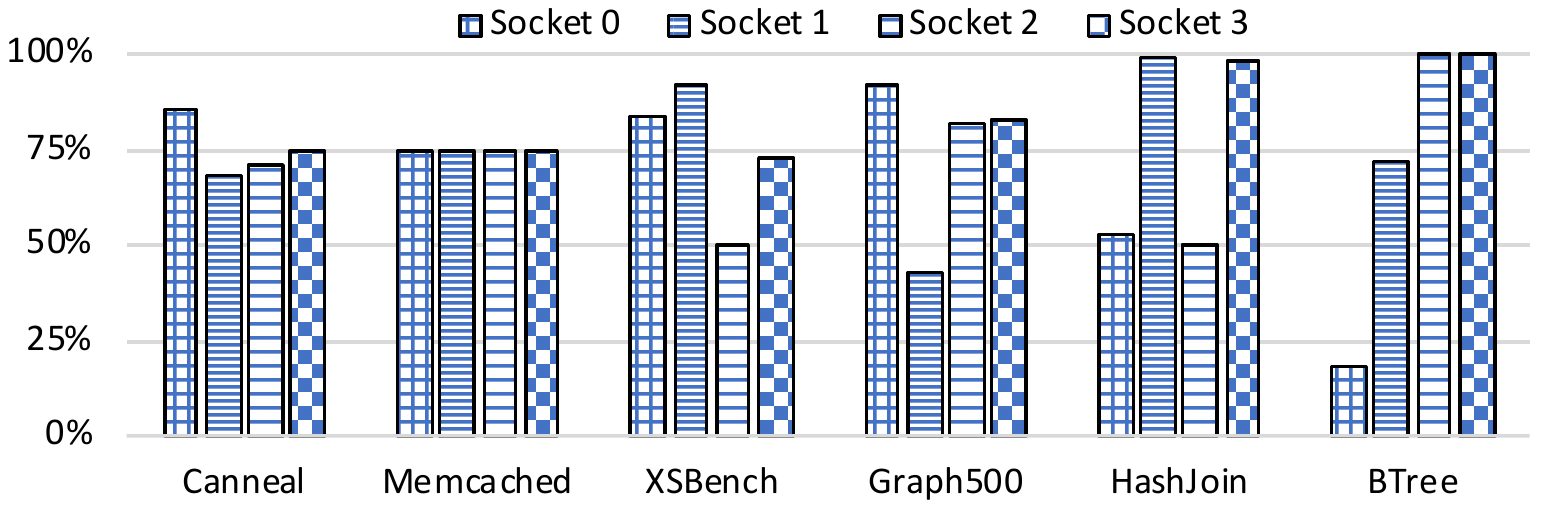}
\end{center}
	\vspace{-0.2cm}
	\caption{Percentage of remote leaf PTEs as observed from each socket for our multi-socket workloads.}
\label{fig:leaf-ptes}
\end{figure}

\noindent
{\bf Results.} Due to space limitations, we show a single, processed snapshot 
of the page-table for Memcached in~\autoref{fig:pgtable:ptrs}. This snapshot
was collected using 4KB pages, local allocation, and AutoNUMA disabled. 
We studied 2MB pages as well and present observations from them later.
The processed dump shows the distribution of all four levels of the page-table (L4 being 
the root, and L1 the leaf). The dump is organized in four columns 
representing the four-sockets in this system. In each cell, the first number is 
the total physical pages at that level-socket combination (e.g. socket 1 
has the only L4 page-table page). Next is the distribution of pointers in square 
brackets of the valid PTEs at this level/socket (e.g. L4 on socket 1 
has 8 pointers to L3 on socket 0, 3 pointers locally, and 1 pointer to socket 
3). The percentage numbers in rounded brackets are the fraction of valid 
PTEs pointing to remote physical pages.

\autoref{fig:leaf-ptes} shows the percentage of remote leaf PTEs observed by a thread running on each socket. 
Each workload's cluster has per-socket values representing the percentage of remote leaf PTEs in the page-table. 
We made these observations from the page-table dumps and distribution of leaf PTEs:

\setlist{nolistsep}
\begin{enumerate}[leftmargin=*,noitemsep]
  \item Page-tables pages are allocated on the socket initializing
    the first data structures that the page-table pages point to. This
    is similar to data frame allocation but has important unintended performance
    consequences. Consider that each page-table page has 512 entries.
    This means that the choice of where to allocate a
    page-table page is entirely dependent upon which of the 512
    entries in the page-table page gets allocated first, and which socket the
    allocating thread runs on. If subsequently, other entries in the
    page-table page are used for threads on another socket,
    remote memory references for page-table walks become common.

  \item With first touch policy, the number of page-tables tends to be
    skewed towards a single socket (e.g. socket 1 for Graph500
    on~\autoref{fig:leaf-ptes}). This is especially the case when a single thread
    allocates and initializes all memory.
  
  \item The interleaved policy evenly distributes page-table pages
    across all sockets.
  
  \item While we observed data pages being migrated with AutoNUMA, page-table 
  pages were never 
  migrated. The fraction of data pages migrated over time depends on the workload and its access locality.
  
  \item On all levels, a significant fraction of page-table entries
  points to remote sockets. In the case of interleave policy, this is 
  $(N-1)/N$ for an $N$-socket system. 
  \item Due to the skew in page-table allocation, some sockets experience longer TLB misses 
	  since up to 99\% of leaf PTEs are located remotely.
\end{enumerate}

\begin{figure*}[t!]
  \centering
  \includegraphics[width=1\textwidth]{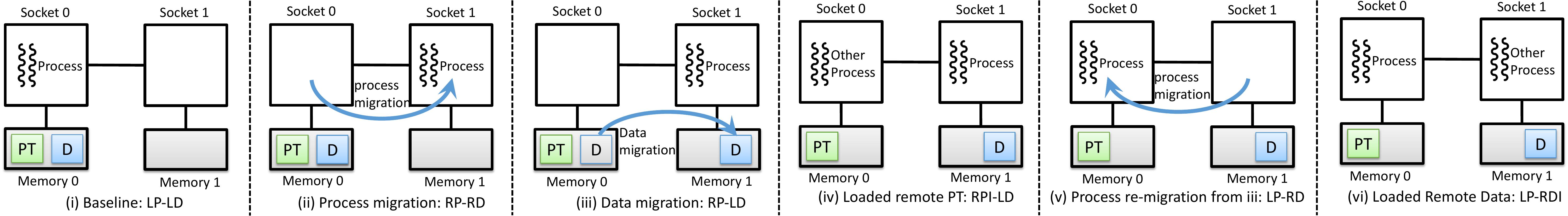}
	\vspace{-0.4cm}
	\caption{Different configurations for workload migration scenario. We show only 6 out of 7 configurations here.
	The 7th configuration (RPI-RDI) can be easily created from (ii) by running another process on Socket 0.}
	\vspace{0.2cm}
	\label{fig:scenarios}
\end{figure*}

\noindent
{\bf Summary} On multi-socket systems, page-table page allocation is
skewed towards sockets that initialize the data
structures. While data pages are migrated by default OS
policies, page-table pages remain on the socket they are
allocated. Consequently, remote page-table walks are inevitable and multi-socket 
workloads suffer from longer TLB misses as their associated page-table walks 
require remote memory accesses.

\subsection{Workload Migration Scenario}
\label{sec:scenario:single}

We now focus on the impact of NUMA on page-table walks in scenarios where
a process on a single socket is migrated to another. Such situations arise 
frequently in commercial cloud deployments due to the need for
 load balancing and improving process-data affinity
\cite{Lozi:2016:LSD, Bouron:2018:BSF}. Particularly, the prevalence
of virtual machines and containers that rely on hypervisors
and NUMA-aware schedulers to consolidate workloads in data centers are making inter-socket process migrations
increasingly common. For e.g., VMware ESXi may migrate 
processes at a frequency of 2 seconds~\cite{vmware:numa}. 
Today, data can be migrated across sockets but page-tables cannot, compromising
performance.

\begin{figure*}[b!]
  \centering
  \includegraphics[width=1\textwidth]{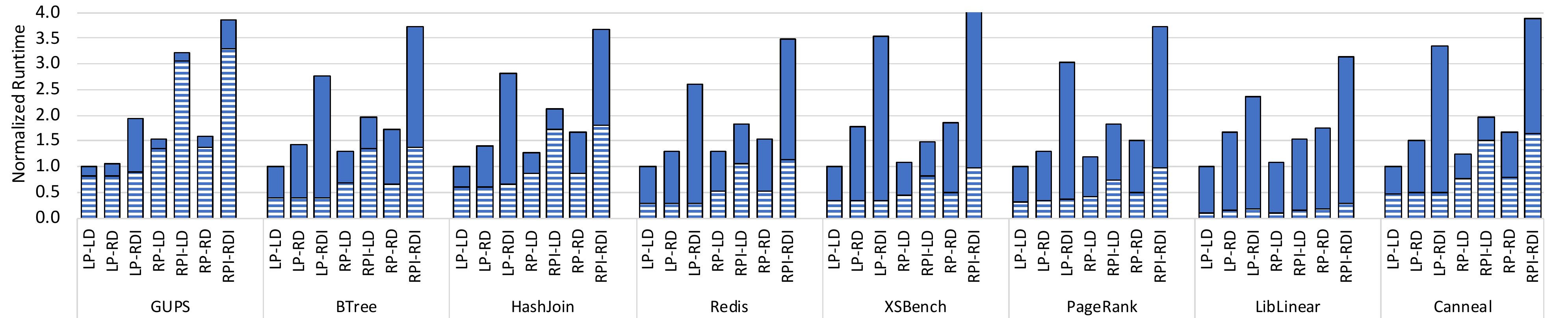}
  \caption{Normalized runtime of our workloads in workload migration scenario with 4KB page
  size. The lower hashed part of each bar is time spent in walking the page-tables. 
	All configurations are shown in~\autoref{tab:scenario:single:config}.}
        \label{fig:analysis}
\end{figure*}

\begin{table}[t!]
  \begin{center}
    \begin{scriptsize}
      \begin{tabular}{lcllll}
              {\bf Config.} & {\bf Workload} & {\bf Page-Table} & {\bf Data} & {\bf Interference} \\ \hline
              (T)LP-LD   &     {A}     &     {A: Local PT}    &      {A: Local Data}     & {-} \\
              (T)LP-RD   &     {A}     &     {A: Local PT}    &      {B: Remote Data}     & {-} \\
              (T)RP-LD   &     {A}     &     {B: Remote PT}    &      {A: Local Data}     & {-} \\
              (T)RP-RD   &     {A}     &     {B: Remote PT}    &      {B: Remote Data}     & {-} \\
              (T)RPI-LD   &     {A}     &     {B: Remote PT}    &      {A: Local Data}     & {B: Interfere on PT} \\
              (T)LP-RDI  &     A     &     {A: Local PT}    &      {B: Remote Data}     & {B: Interfere on Data} \\
              (T)RPI-RDI &     A     &     {B: Remote PT}    &      {B: Remote Data}     & {B: Interfere on PT\&Data} \\
      \end{tabular}
    \end{scriptsize}
  \end{center}
        \vspace{-0.2cm}
  \caption{Configurations for workload migration scenario, where A and B
        denote different sockets. T denotes if THP in Linux is used for 2MB pages.
        Interference is another process that runs on a specified socket and hogs its local memory bandwidth.
        ~\autoref{fig:scenarios} shows the 2-socket case.}
        \vspace{-0.2cm}
        \label{tab:scenario:single:config}
\end{table}

\noindent
{\bf Configurations.} We run each workload in isolation while tightly
controlling and changing \ei the allocation policies for data pages
and page-table pages, \eii whether or not the sockets are idle and
\eiii whether transparent, 2MB large pages (THP) are enabled. We
disable NUMA migration. To study page-table allocations in a
controlled manner, we modified Linux kernel to force page-table
allocations on a fixed socket. We use the
configurations shown in~\autoref{tab:scenario:single:config} and
visualized in~\autoref{fig:scenarios}.  We use the STREAM
benchmark~\cite{stream} running on the socket indicated by interference to
create a worst-case scenario of co-locating a memory-bandwidth heavy workload. 
Memory allocation and processor affinity are controlled by \texttt{numactl}.

\noindent
{\bf Measurements.} We use \texttt{perf} to obtain performance
counter values such as execution cycles and TLB load and store miss
walk cycles (i.e., the cycles that the page walker is active
for).

\noindent
{\bf Results.} We then run our workloads for all seven configurations.
~\autoref{fig:analysis} shows the normalized run times with a 4KB page size.
The base case is the LP-LD configuration where both page-tables
and data pages are local and the system is idle. For each
configuration, hashed part of the bar denotes the fraction of time spent
on page-table walks. We observe the following from this experiment:

\setlist{nolistsep}
\begin{enumerate}[leftmargin=*,noitemsep]
  \item All workloads spend a significant fraction of execution cycles
    (up to 90\%) performing page-table walks. Parts of these walks may
    be overlapped with other work; nevertheless, they present a
    performance impediment.
  \item LP-LD runs most efficiently for 4KB page size.
  \item The local page-table, remote data case (LP-RD and LP-RDI)
    suffers 3x slowdown versus the baseline. This
    is not surprising and has motivated prior research on
    data migration techniques in large-memory NUMA machines.
  \item More surprisingly, the remote page-table, local data case
    (RP-LD and RPI-LD) suffers 3.3x slowdown. This slowdown can 
    even be more severe than remote data accesses.
  \item When both page-tables and data pages are placed remotely (RP-RD 
  and RPI-RDI), the slowdown is 3.6x and is the worst 
  placement possible for all workloads.
  \item With 2MB page size (figure omitted for space), TLB reach improves and the number of
    memory accesses for a page-table walk decreases to 3 rather than
    4. These two factors reduce the fraction of execution cycles
    devoted to page-table walks. Even so, overall performance
    is still vulnerable to remote page-table placement.
\end{enumerate}
{\bf Summary.} The NUMA node on which page-table pages are placed
significantly impacts performance. Remote page-tables can have similar, 
and in some cases even worse, slowdown
than remote data pages accesses. Moreover, the
slowdown is visible even with large pages.

\begin{figure*}[t]
  \centering
  \includegraphics[width=\textwidth]{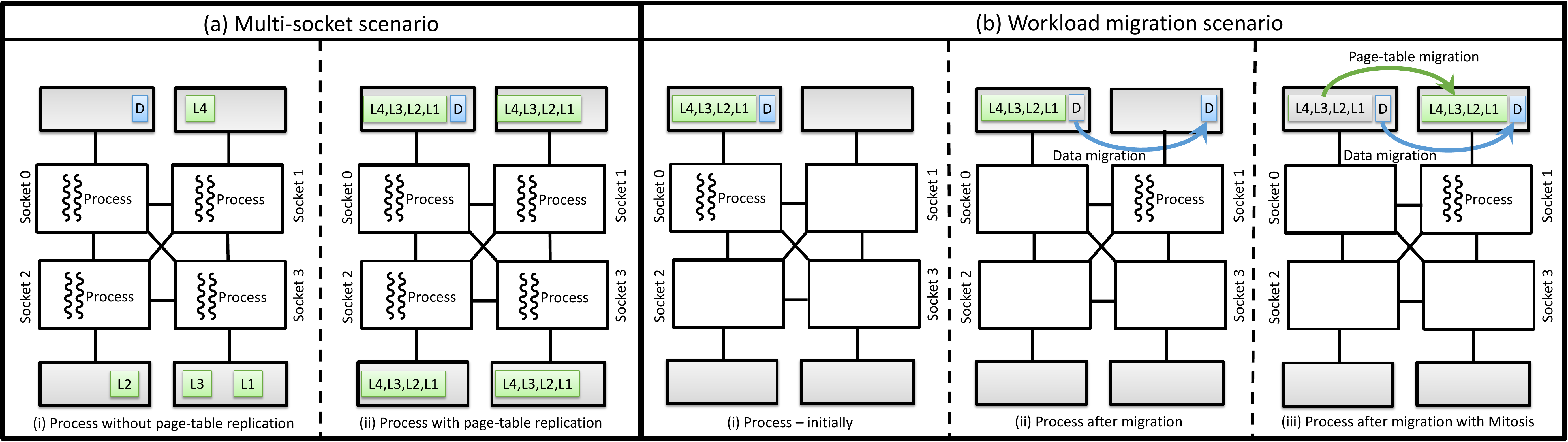}
  \caption{\sys: Page-table migration and replication on large-memory machines}
  \label{fig:migration}
\end{figure*}

%% file: 40-design.tex
\section{Design Concept}
\label{sec:design:concept}

\sys' key concept is a mechanism and its policies to replicate and migrate page-tables
and reduce the frequency of remote memory accesses in page-table walks. 
\sys requires two components: \ei a mechanism to support low-overhead 
page-table replication and migration and \eii policies for processes to 
efficiently manage and control page-table replication and migration.
~\autoref{fig:migration} illustrates these concepts. Our discussion
focuses on the multi-socket and workload migration 
scenarios used before in \autoref{sec:overheads}.

\subsection{Multi-socket Scenario}
We showed in~\autoref{sec:scenario:multi} that multi-socket workloads will,
assuming a uniform distribution of page-table pages, have
$\frac{N-1}{N}$ PTEs pointing to remote pages for an $N$-socket
system. Page-tables may be distributed among the sockets in
a skewed fashion.~\autoref{fig:migration} (a)(i) shows a scenario 
where threads of the same workload running on different sockets have 
to make remote memory accesses during page-table walks.

From~\autoref{fig:migration} (a)(i) we can see that if a thread in
socket 0 has a TLB miss for data ``D'' (which is local to the socket), it
has to perform up to 4 remote accesses to resolve the TLB miss to only
find out that the data was local to its socket.

With \sys, we replicate the page-tables on each socket where the
process is running (shown in~\autoref{fig:migration} (a)(ii)). This
results in up to 4 local accesses to the page-table, precluding the need
for remote memory accesses in page-table walks. 

\subsection{Workload Migration Scenario}
Single-socket workloads suffer performance loss when processes
are migrated across sockets while page-tables are not (shown in ~\autoref{fig:migration} (b)(ii)) 
The process is migrated from socket 0 to socket 1,
the NUMA memory manager transparently migrate data pages, but 
page-table pages remain on socket 1. In contrast, \sys migrates the page-tables 
along with the data (\autoref{fig:migration} (b)(iii)). This 
eliminates remote memory accesses for page-table walks, improving performance.

\section{Mechanism}
\label{sec:design:mechanism}

Replication and migration are inherently
similar. We first describe the building blocks which are required
to support page-table replication and later show how we can leverage the
replication infrastructure to achieve page-table migration.

\sys enables per-process replication; the virtual memory subsystem
needs to maintain multiple copies of page-tables for a single process. 
Efficient replication of page-tables can be divided into three
sub-tasks: \ei strict memory allocation to hold the replicated
page-tables, \eii managing and keeping the replicas consistent, and
\eiii using replicas when the process is scheduled. We now describe
each sub-task in detail by providing a generalized design and our Linux implementation.
We also discuss how \sys handles accessed and dirty bits.

\subsection{Allocating Memory for Storing Replicas}  

{\bf General design:} 
All page-table allocations are performed by the OS on a page-fault--an explicit
mapping request can be viewed as an eager call to the page-fault handler
for the given memory area. \sys extends the same mechanism to
allocate memory across sockets for different replicas.

Such allocation is strict, i.e. it has to occur on a
particular list of sockets at allocation time. It is, therefore, possible
that it may fail due to the unavailability of memory on those
sockets. There are multiple ways to sidestep this problem by reserving pages on 
each socket for page-table allocations using per-socket \emph{page-cache}. These 
pages can be explicitly reserved through a system call or automatically when a 
process allocates a virtual memory region. Alternatively, the OS can reclaim 
physical memory through demand paging mechanisms or evicting a data page onto
another socket.

\noindent
{\bf Linux implementation:} 
We rely on the existing page allocation functionality in Linux to implement
\sys. When allocating page-table pages, we explicitly supply the list of 
target sockets for page-table replication.
Since strict allocation can fail, we implemented per-socket \emph{page-caches} 
to reserve pages for page-table allocations. The size of this
page-cache is explicitly controlled using a \texttt{sysctl} interface.

\subsection{Management of Updates to Replicas}  

{\bf General design:} 
For security, OSes usually do not allow user processes to directly
manage their own page-tables. Instead, OSes export an interface
through which page-table modifications are handled,
e.g.\ map/unmap/protect of pages. 
\sys extends the same interfaces for updates to page-tables to keep all replicas 
consistent. One way to implement this is to \emph{eagerly} update all replicas 
at the same time via this standard interface when an update to the 
page-table is performed on any replica. 

\begin{figure}[t]
  \centering
  \includegraphics[width=\columnwidth]{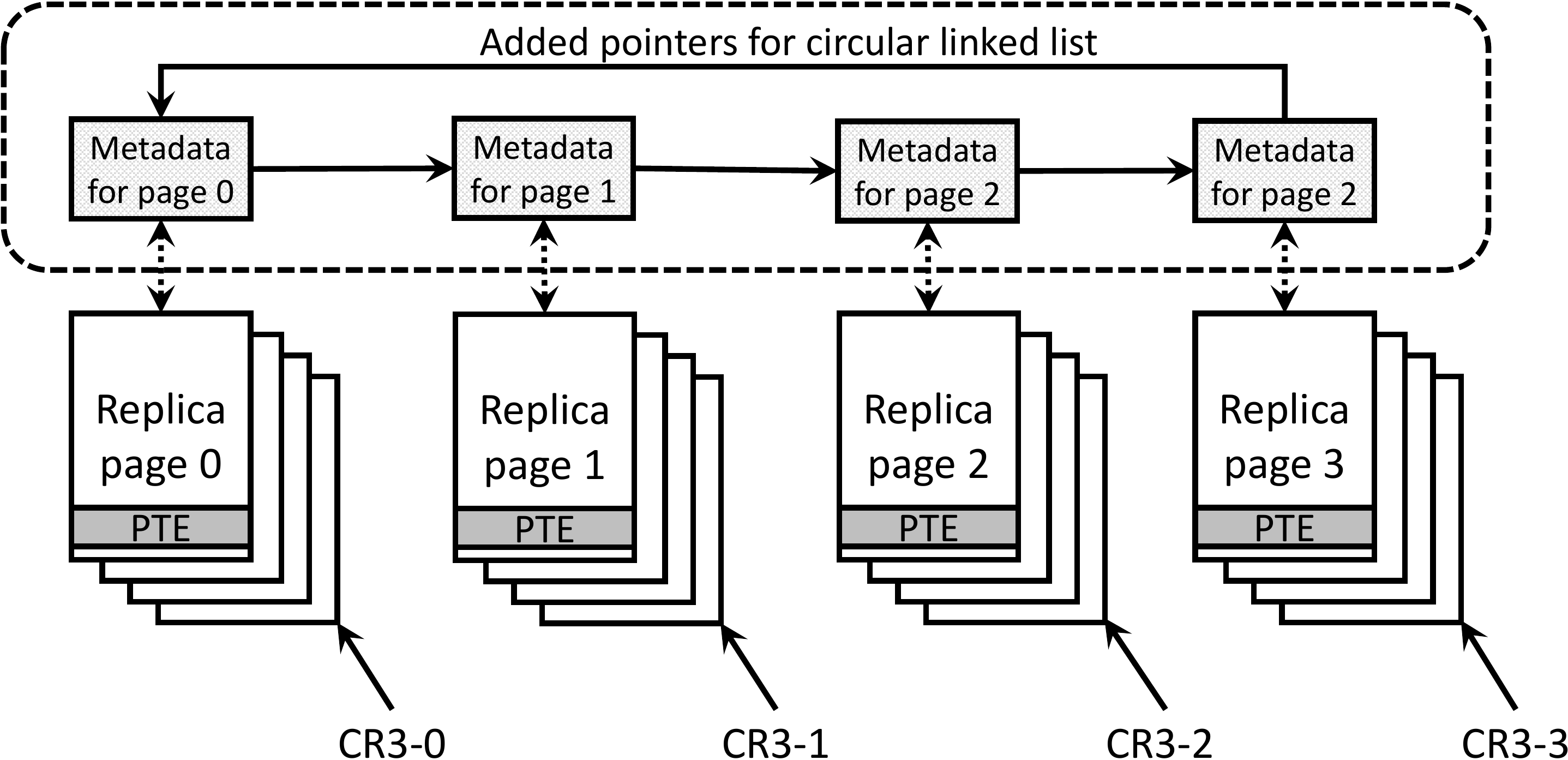}
        \caption{Circular linked list to locate all replicas efficiently (implemented in Linux with \codeinline{struct page}).}
  \label{fig:circular}
\end{figure}

On an eager update, the OS finds the physical location to update in the local replica by walking the local replica of the page-table.
It is required to walk other replicas of the page-table to locate the physical location to update all the replicas at the same time. 
Therefore, an N-socket system in x86\_64 will need $4N$ memory accesses with 
replication on a page-table update: 
4 memory accesses to walk the page-table on each of the N sockets.
To reduce this overhead, we designed a circular linked-list of all replicas. The metadata about each physical page
is utilized to store the pointers to the next physical page holding the replica of the page-table. 
\autoref{fig:circular} shows an illustration with 4-way replication. 
This allows updates to proceed without walking the page-tables to perform the update.
With this optimization, the update of all $N$ replicas takes $2N$ memory 
references ($N$ for updating the $N$ replicas and $N$ for reading the pointers 
to the next replica).

\noindent
{\bf Linux implementation:}
We implemented eager updates to the replica page-tables in Linux. 
This required intercepting any writes to the page-tables and propagate updates accordingly.
But instead of revamping the full-memory subsystem in Linux, we used a different interface,
PV-Ops~\cite{pvops}, which is required to support para-virtualization environments such as Xen~\cite{Barham:2003:XAV}.
The Linux kernel shipped with distributions like Ubuntu has para-virtualization
support enabled by default.

Conceptually, this is done by indirect calls to the native or Xen handler
functions. Effectively, the indirect calls are patched with direct calls once
the subsystem is initialized. The PV-Ops subsystem interface consists of
functions to allocate and free page-tables of any level, reading and writing
the translation base register (CR3 on x86\_64), and writing page-table entries.
The PV-Ops interface can be seen in~\autoref{lst:pvops}.

\begin{lstlisting}[language=C, style=CStyle,label=lst:pvops,caption={Excerpt of 
the PV-Ops interface}] 
void write_cr3(unsigned long x);
void paravirt_alloc_pte(struct mm_struct *mm, unsigned long pfn);
void paravirt_release_pte(unsigned long pfn);
void set_pte(pte_t *ptep, pte_t pte);
\end{lstlisting}

We implemented \sys as a new backend for PV-Ops alongside with the native
and Xen backends.
When the kernel is compiled with \sys, the default PV-Ops is
switched to the \sys backend. We implemented the \sys backend with great care
to ensure identical behavior to the native backend when \sys is turned off.
Besides, note that replication is generally not enabled by default, and
thus the behavior is the same as the native interface.

The PV-Ops subsystem provides an efficient
way for \sys to track any writes to the page-tables in the system.
Propagating those updates efficiently requires a fast way to find the replica
page-tables based solely on the information provided through the PV-Ops
interface (\autoref{lst:pvops}) i.e.\ using a kernel virtual address (KVA)
or a physical frame number (PFN).

We augment the page metadata to keep track of replicas with our circular linked list. 
The Linux kernel keeps track of each 4KB physical frame in the system using 
\codeinline{struct page}. Moreover, each frame has a unique KVA and PFN. 
Linux provides functions to convert between \codeinline{struct page} and it's 
corresponding KVA/PFN, which is typically done by adding, subtracting or 
shifting the respective values and are hence efficient operations.
We can, therefore, obtain the \codeinline{struct page} directly from the 
information passed through the PV-Ops interface and update all replicas efficiently.

\subsection{Efficiently Utilizing Page-Table Replicas} 

{\bf General design:} When the OS schedules a process or task, 
it performs a context switch, restores processor
registers and resumes execution of the new process or task. The
context switch involves programming the page-table base register of
the MMU with the base address of the process' page-table and
flushing the TLB. With \sys, we extend the context switch functionality, 
to select and set the base address of the socket's local
page-table replica efficiently. This enables a task or process to
use the local page-table replica if present.

\noindent
{\bf Linux implementation:} For each process, we maintain an array of root page-table pointers 
which allows directly selecting the local replica by indexing this array using the socket id. 
Initializing this array with pointers to the very same root page-table is equivalent to the native behavior.

\subsection{Handling of Bits Written by Hardware}

{\bf General design:} A page-table is mostly managed by software (the OS) most of the time and read by 
the hardware (on a TLB miss). On x86, however, hardware--namely the 
page-walker--reports whenever a page has been accessed or written to by setting 
the accessed and dirty bits in the PTEs. In other words, 
page-table is modified without direct OS involvement. Thus, accessed and dirty 
bits do not use the standard software interface to update the PTE and cannot be 
replicated easily without hardware support. Note, that these two bits
are typically set by the hardware and reset by the OS. They are used by the 
OS for system-level operations like swapping or writing back memory-mapped 
files if they are modified in memory.  With \sys when replicated, we 
logically OR accessed and dirty bits of all the replicas when read by the 
OS.

\noindent
{\bf Linux implementation:} 
We need to read accessed/dirty bits from all replicas as well as reset them in all replicas. 
Unfortunately, the PV-Ops interface doesn’t provide functions
to read a page-table entry, worse we have found code
in the Linux kernel which even writes to the page-table entry
without going through the PV-Ops interface. We augmented
with the corresponding \codeinline{get} functions to PV-Ops which consult
all copies of page-table entry and make sure the flags are
returned correctly. The new function reads all the replicas
and ORs the bits in all replicas to get the correct information.

\subsection{Page-Table Migration} 
We use replication to perform migration in the 
following way: we use \sys to replicate the page-table on the socket
to which the process has been migrated. The first replica can be eagerly freed after 
migration, or alternatively kept up-to-date in the case the process gets 
migrated back and lazily deallocated in case physical memory is becoming scarce.

\section{Policy}
\label{sec:design:policy}

The policies we implement with \sys control when page-tables are
replicated and determine the processes and sockets for which
replicas are created. As with NUMA policies, page-table replication
policies can be applied system-wide or upon user request. We discuss
both in this section.

\subsection{System-wide Policies} 

{\bf General design:} System-wide policies can range from simple
on/off knobs for all processes to policies that actively monitor
performance counter events provided by the hardware to dynamically 
enable or disable \sys.

Event-based triggers can be developed for page-table migration and
replication within the OS. For instance, the OS can obtain TLB miss
rates or cycles spent walking page-tables through performance counters
that are available on modern processors and then apply policy
decisions automatically. A high TLB miss rate suggests that a
process can benefit from page-table replication or
migration. By taking the ratio between the time spent to serve TLB
misses and the number of TLB misses can indicate a
replication candidate. Processes with a low TLB miss rate may not
benefit from replication.

Even if the OS makes a decision to migrate or replicate the
page-tables, there it may be costly to copy the entire page-table as big
memory workloads easily achieve page-tables of multiple GB in size. By
using additional threads or even DMA engines on modern processors, the
creation of a replica can happen in the background and the application
regains full performance when the replica or migration has completed.

The target applications of \sys are long-running, big-memory workloads with 
high TLB pressure, and therefore we disable page-table replication
for short-running processes since the performance and memory cost of
the replicated page-tables for short-running processes cannot be amortized
(\autoref{sec:eval:memoryoverheads}).

\noindent
{\bf Linux implementation:}
We support a straightforward, system-wide policy with four states: \ei  
completely disable \sys, \eii enable per-process basis, \eiii fix the 
allocation of page-tables on a particular socket, and \eiv enabled for all 
processes in the system. This system-wide policy can be set through the 
\texttt{sysctl} interface of Linux. 
We leave it as future work to implement an automatic, counter-based 
approach.

\subsection{User-controlled Policies}  

{\bf General design:} System-wide policies usually imply a
one-size-fits-all approach for all processes, but user-controlled
policies allow programmers to use their understanding of their
workloads and to select policies explicitly. These user-defined
replication and migration policies can be combined with data and
process placement primitives. Such policies can be selected when
starting the program by defining the CPU set and replication set, or
at runtime using corresponding system calls to set affinities and
replication policies. All of these policies can be set per-process so
that users have fine-grained control on replication and migration.

\noindent
{\bf Linux implementation:} We implement user-defined policies as an additional 
API call to \texttt{libnuma} and corresponding parameters of \texttt{numactl}. Similar 
to setting the allocation policy, we can supply node-mask or a list of sockets 
to replicate the page-tables (\autoref{lst:numactl}). Applications can 
thus select the replication policy at runtime, or we can use 
\texttt{numactl} to select the policy without changing the program.

\begin{lstlisting}[language=C, 
style=CStyle,label=lst:numactl,caption={Additions to libnuma and 
numactl}]
numactl [--pgtablerepl= | -r <sockets>]
void numa_set_pgtable_replication_mask(struct bitmask *);
\end{lstlisting}

Both, \texttt{libnuma} and \texttt{numactl} use two additional system calls to set 
and get the page-table replication bitmask. 
Whenever a new mask is set, \sys will 
walk the existing page-table and create replicas according to the new bitmask. 
The bitmask effectively specifies the replication factor: $N$ bits set 
corresponds to copies on $N$ sockets and by passing an empty bitmask, the default 
behavior is restored.

%% file: 51-discussion.tex
\section{Discussion}
\label{sec:discussion}

\subsection{Why Linux Implementation?}

As a proof-of-concept, we implement \sys
in the widely-used Linux OS. Choosing Linux as our testbed allows us
to prototype our ideas on a complex and complete OS where the subtle
interactions of many systems features and \sys stress-tests its
evaluation. Specifically, we use mainline Linux kernel v4.17 and
implement \sys for the x86\_64 architecture. We plan to release
this implementation for everyone to use and plan to upstream the 
changes to the Linux kernel.

\subsection{Applicability to Library OS}

We have chosen to implement the prototype of \sys in Linux. However, the concept 
of \sys is applicable to other operating systems. Microkernels, for instance, 
push most of their memory management functionality into user-space
libraries or processes while the kernel enforces security and
isolation.  In Barrelfish~\cite{Baumann:2009}, for example, processes
manage their own address space by explicit capability invocations to
update page-tables with new mappings.

In such a system, one could implement \sys purely in user-space by linking 
to a \sys-enabled libraryOS, and the kernel itself would not need 
to be modified at all. The library can keep track of the address space, 
including page-tables, replicas etc.  Those data-structures can easily be 
enhanced to include an array of page-table capabilities instead of a
single such table. This would allow policies to be defined at
application level by using an appropriate policy library.  Updates to
page-tables might need to be converted to explicit update messages to
other sockets, which avoid the need for global locks and propagates updates 
lazily.  On a page-fault, updates can be processed and applied
accordingly in the page-fault handling routine.  We leave such an
implementation to future work, but believe it to be straightforward.

\subsection{Huge/Large Pages Support?}

Larger page sizes help reduce address translation overheads by increasing 
the amount of memory that each TLB entry map by orders of magnitude. Even with 
2MB and 1GB page size support in x86-64 on an Intel Haswell processor, 
the TLB reach is still less than 1\%, assuming 1TB of main memory for any page 
size. Moreover, many commodity processors provide limited numbers of large page 
TLB entries especially 1GB TLB entries, which limits their benefit 
~\cite{Basu:2013:dir_seg,Gandhi:2014:emv, Karakostas:2014} and additionallly 
huge-pages are not always the best choice~\cite{Gaud:2014:LPM}. 

Since, address translation overheads are non-negligible with larger page sizes,
they are susceptible to NUMA effects on page-table walks. Thus, our 
implementation of \sys supports larger page sizes and evaluate them. 
We extend transparent huge pages (THP) or 2MB page size in Linux which requires 
coalescing smaller pages to a large page and splitting larger pages in to 
smaller ones. 
\sys is implemented to replicate the page-tables even in presence of such 
mechanisms.

\subsection{Applicability to Virtualized Systems?}
Virtualized systems widely use
hardware-based nested paging to virtualize memory~\cite{Gandhi:2017:micro}. This requires
two-levels of page-table translation:
\begin{enumerate}[leftmargin=*,noitemsep]
\item gVA to gPA: guest virtual address to guest physical address
     via a per-process guest OS page-table (gPT)
\item gPA to hPA: guest physical address to host physical address via a per-VM
     nested page-table (nPT)
\end{enumerate}
In the best case, the virtualized address translation hits in the TLB to directly
translate from gVA to hPA with no overheads. In the worst case, a TLB miss needs
to perform a 2D page walk that multiplies overheads vis-a-vis native, because
accesses to the guest page-table also require translation by the nested page-table.
For x86-64, a nested page-table walk requires up to 24 memory accesses. This 2D
page-table walk comes with additional hardware complexity.

Understanding page-table placement in virtualized systems is a major 
undertaking and requires a separate study.  
We believe we can extend \sys' design to replicate both guest 
page-tables and nested page-tables independently if the underlying NUMA 
architecture is exposed to the guest OS to improve performance of applications. 
To extend the design, we can rely on setting accessed and dirty bits at 
both gPT and nPT by the nested page-table walk hardware available since Haswell~\cite{intel:eptadbits}. 
Thus, we can extend our OS extension for or-ing the access and dirty bits 
across replicas to get the correct information at both levels independently.
However, the main issue is that most cloud systems prefer not to expose the underlying architecture to
the guest OS making a case for novel approaches to replicate and migrate both
levels of page-tables in a virtualized environment.

\subsection{Consistency across page-table replicas?}

Coherence between hardware TLBs is maintained
by the OS with the help of TLB flush IPIs and updates to the page-table are already thread-safe 
as they are performed within a critical section. In Linux, a lock is taken
whenever the page-table of a process is modified and thus ensuring mutual
exclusion. The updates to the page-table structure are made visible after
releasing the lock. When an entry is modified, its effect is made visible to
other cores through a global TLB flush as the old entry might still be cached.

With \sys, we currently keep the same consistency guarantees 
by updating
all page-table replicas eagerly while being in the critical section. Thus, only
one thread can modify the page-table at a time. Hardware may read the page-table
while updates are being carried out. The critical section ensures correctness
while serving the page-fault while again, the global TLB flush ensures consistency
after modification of an entry in case a core has cached the old one.

%% file: 60-evaluation.tex
\section{Evaluation}
\label{sec:evaluation}

\begin{figure*}[t]
  \centering
  \begin{subfigure}{.5\textwidth}
    \centering
    \includegraphics[width=1\textwidth]{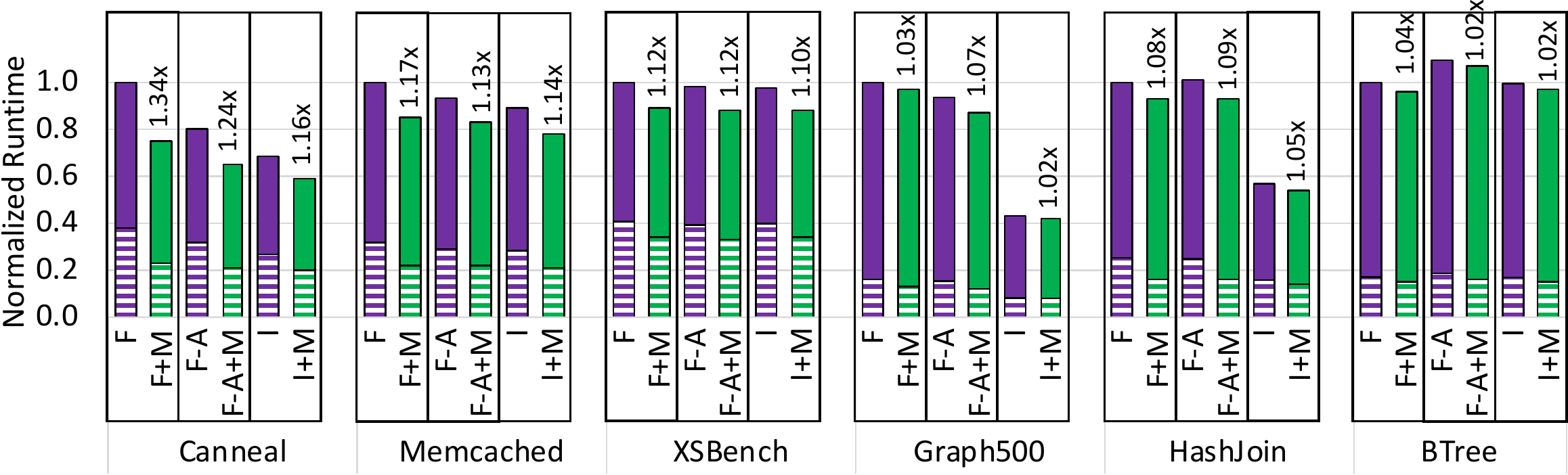}
    \caption{4KB Pages}
    \label{fig:eval:multi:mitosis}
  \end{subfigure}%
  \begin{subfigure}{.5\textwidth}
    \centering
    \includegraphics[width=1\textwidth]{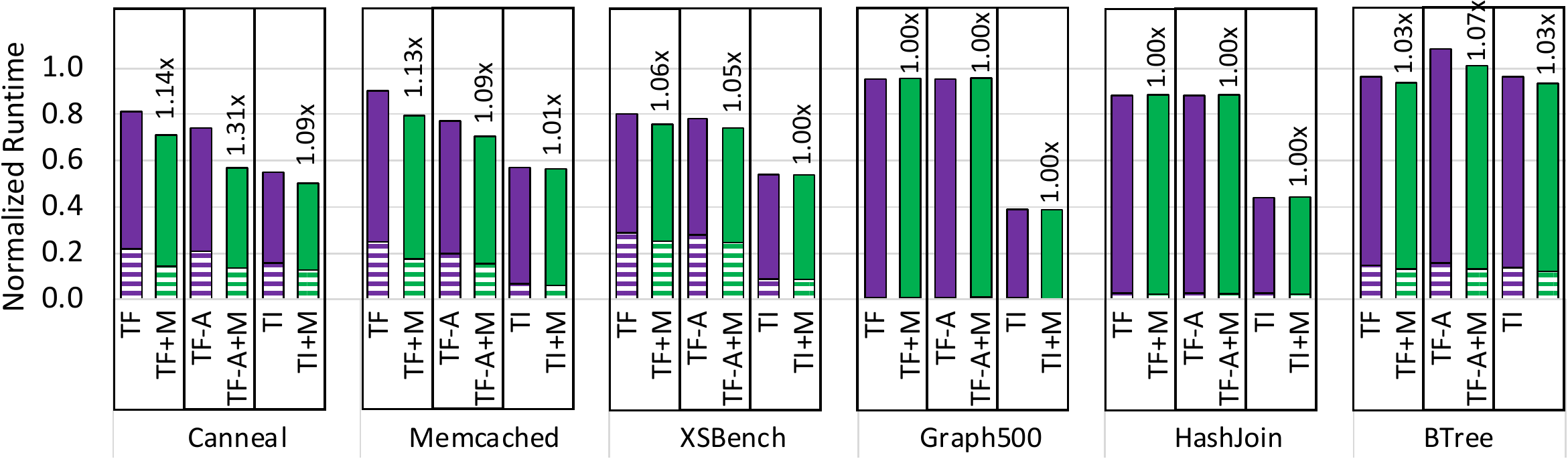}
    \caption{2MB Large Pages}
    \label{fig:eval:multi2m:mitosis}
  \end{subfigure}
 \vspace{-0.2cm}
  \caption{Normalized performance with \sys for multi-socket
  workloads with 4KB and 2MB page
        size.
    The lower hashed part of each bar is execution time spent in walking the page-tables.
        }

\end{figure*}

\begin{figure*}[b]
  \centering
   \begin{subfigure}{.5\textwidth}
    \centering
    \includegraphics[width=1\textwidth]{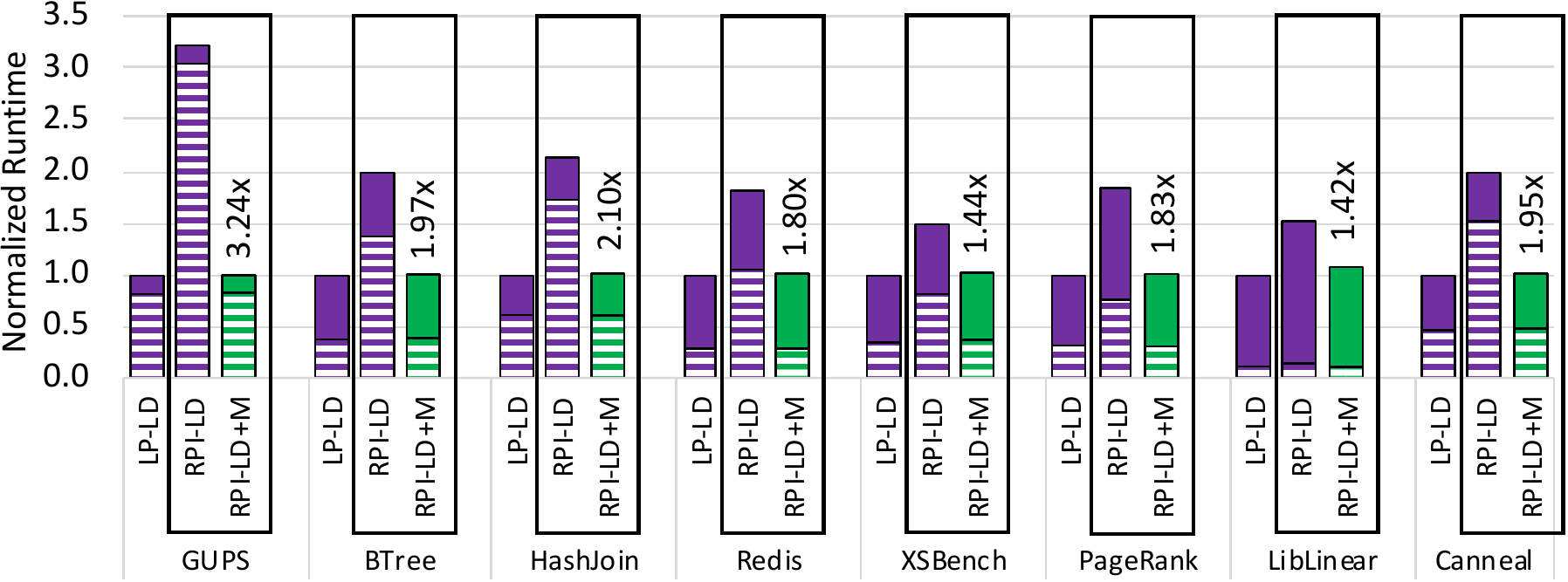}
    \caption{4KB Pages}
    \label{fig:eval:single:mitosis}
  \end{subfigure}%
  \begin{subfigure}{.5\textwidth}
    \centering
    \includegraphics[width=1\textwidth]{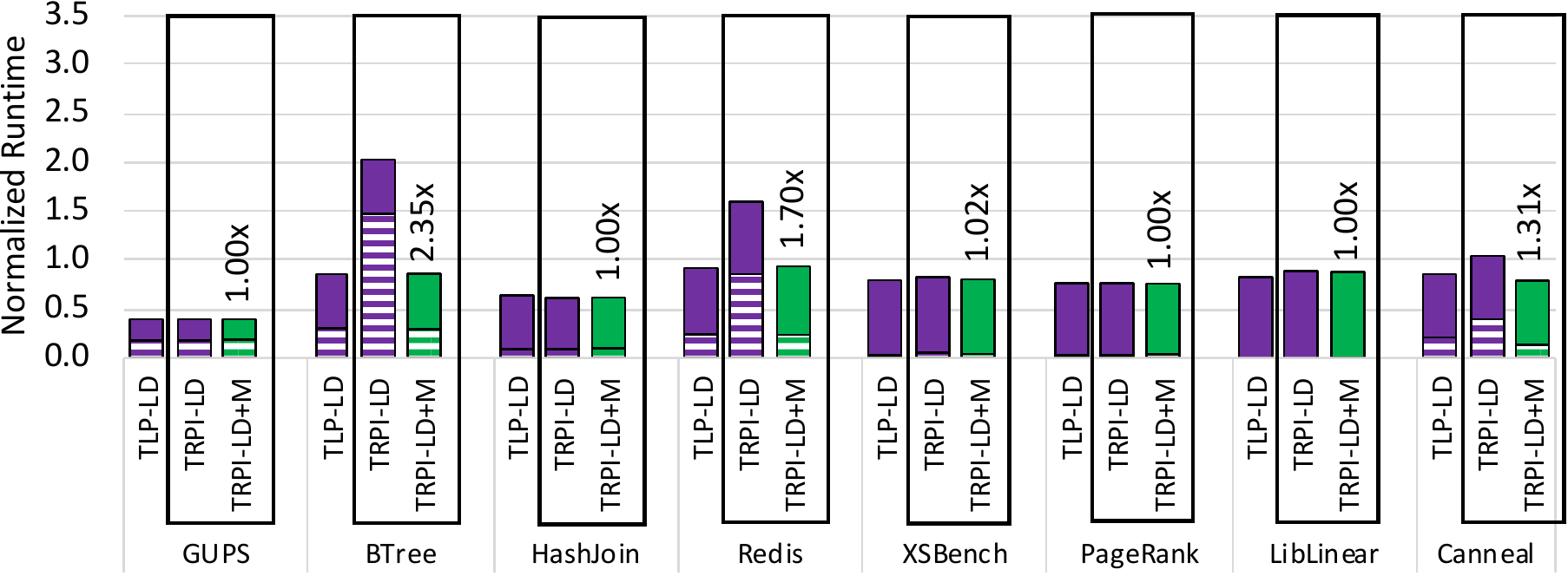}
    \caption{2MB Large Pages}
    \label{fig:eval:single2m:mitosis}
  \end{subfigure}
	\vspace{-0.2cm}
  \caption{Normalized performance with \sys for workloads in workload 
  migration scenario with 4KB 
	and 2MB page size. 
	The lower hashed part of each bar is execution time spent in walking the page-tables.}
\end{figure*}

We evaluate \sys using a set of big-memory workloads and micro-benchmarks. 
We show: (1) how multi-threaded programs benefit from \sys (\autoref{sec:eval:multithread}),
(2) how \sys eliminates NUMA effects of page-walks when page-tables
are placed on remote sockets due to task migration (\autoref{sec:eval:singlethread}) and 
(3), the memory and runtime overheads of \sys (\autoref{sec:eval:memoryoverheads}).

\paragraph{Hardware Configuration} 
\label{sec:eval:hardware}
We used a four-socket Intel Xeon E7-4850v3 with 14 cores and 128GB memory per-socket (512 GB total memory) with 2-way 
hyper-threading running at 2.20GHz. 
The L3 cache is 35MB in size and the processor has a per-core two-level TLB with 64+1024 
entries. Accessing memory on the local NUMA socket has about 
280 cycles latency and throughput of 28GB/s. For a remote NUMA socket, this is 
580 cycles and 11GB/s respectively.

\subsection{Multi-socket Scenario}
\label{sec:eval:multithread}

In this part of the evaluation, we focus on multi-threaded workloads running in 
parallel on all sockets in the system. For a machine with $N$ NUMA sockets, in 
expectation $\frac{N-1}{N}$ of page-table accesses will be remote while the 
remote sockets are busy themselves.
We evaluate six workloads (see \autoref{sec:scenario:multi}),
for all commonly used configurations that influence data
and page-table placement (see ~\autoref{tab:scenario:multi:config}). Performance is presented as an average of three runs, excluding the initialization phase.

The results are shown in~\autoref{fig:eval:multi:mitosis} for 4KB pages and~\autoref{fig:eval:multi2m:mitosis} with 2MB large pages respectively.
All bars are normalized to 4KB first-touch allocation policy (bar: F).
Bars with the same allocation policy are grouped in boxes for comparison.
The number on top of \sys bars (green) shows improvement from corresponding non-\sys bars (purple) within a box.
Note that data allocation policy impacts performance and is shown across boxes for each workload.
The results for 2MB pages are normalized to 4KB (bar: F) to show performance
impact with increase in page size.

\begin{table}[t]
  \begin{center}
    \begin{footnotesize}
        {\renewcommand{\arraystretch}{1.2}%
            \begin{tabular}{lll}
                {\bf Config.} & {\bf Data pages} & {\bf Page-table pages}\\ \hline
                    {(T)F}              &     \multirow{2}{*}{First-touch allocation}                           & {First-touch allocation (bar: purple)} \\ \cline{1-1}\cline{3-3}
                    {(T)F+M}            &                               & {\emph{Mitosis replication (bar: green)}} \\ \hline
                     {(T)F-A} &     {First-touch allocation}  & {First-touch allocation (bar: purple)}   \\ \cline{1-1} \cline{3-3}
                     {(T)F-A+M}&     {+ Auto page migration}    & {\emph{Mitosis replication (bar: green)}}\\ \hline
                     {(T)I}     &    \multirow{2}{*}{Interleaved allocation}                            & {Interleaved allocation (bar: purple)}  \\ \cline{1-1} \cline{3-3}
                     {(T)I+M}  &                                & {\emph{Mitosis replication (bar:green)}}     \\ \hline
            \end{tabular}
            }
    \end{footnotesize}
  \end{center}
        \vspace{-0.4cm}
  \caption{Configurations for multi-socket scenario where workload runs on all sockets.
        T denotes Linux with THP. M denotes the corresponding data allocation policy with \sys.}
        \vspace{-0.3cm}
  \label{tab:scenario:multi:config}
\end{table}

We observe that with 4KB pages, up 
to 40\% of the total runtime is spent in servicing TLB misses. 
\sys reduces the overall runtime for all applications with the best-case improvement of 1.34x for Canneal.
Most of the improvements can be noted in the reduction of page-walk cycles due to replication of page-tables.

Large pages can significantly reduce translation overheads for many workloads. However,
NUMA effects of page-table walks are still noticeable, even if all workload memory is backed by large pages. Hence, \sys provides significant speedup e.g.,  1.14x, 1.13x, 1.06x and 1.07x for Canneal, Memcached, XSBench and BTree, respectively.
Note that the use of  large pages can lead to decreased performance on NUMA 
systems and still not used for many systems~\cite{Gaud:2014:LPM}. 

Using various data page placement policies improves performance for 
our workloads as expected. In combination with all policies, \sys 
consistently improves performance.

We have provided evidence that highly parallel workloads experience NUMA effects 
of remote-memory accesses due to page-table walks. Yet, running a workload 
concurrently means we cannot inspect a thread in isolation: a TLB miss on one 
core may populate the cache with the PTE needed to serve the 
TLB miss on another core of the same socket. Moreover, accessing a remote last-level 
cache may be faster than accessing DRAM. Nevertheless, we have shown 
that \sys is still able to improve multi-threaded workloads by up to 1.34x and that too for both page sizes. 
Again, \sys does not cause any slowdown.

\subsection{Workload Migration Scenario}
\label{sec:eval:singlethread}


As we observed in \autoref{sec:scenario:single}, NUMA
schedulers can move processes from one socket to another under various constraints.
In this part of the evaluation, we show that \sys eliminates NUMA
effects of page-walks originating due to data and threads migrating to a different socket while page-tables remain
fixed on the socket where workload was first initialized.

We execute the same workloads used for workload migration
scenario in \autoref{sec:scenario:single}. As an additional configuration, we
enabled \sys when the page-table is allocated on a remote socket. Recall, we
disabled Linux' AutoNUMA migration, and pre-allocated and initialized the
working set (17-85GB).

The results are shown in~\autoref{fig:eval:single:mitosis}
and~\autoref{fig:eval:single2m:mitosis} with 4KB and 2MB page sizes
respectively.~\autoref{tab:scenario:single:config} in~\autoref{sec:scenario:single} showed the configurations used for evaluation: LP-LD (Local PT - Local Data) and RPI-LD (Remote PT with interference - Local Data).
RPI-LD+M shows the improvement with page-table migration enabled by \sys when RPI-LD case arises in the system.
The boxes denote the bars to compare to see the improvement due to page-table migration. The number on top of the
bar denotes the improvement due to \sys (green bar) as compared to non-mitosis bar (purple bar) within the same box.
All bars are normalized to 4KB LP-LD configuration.
The results for 2MB pages are normalized to 4KB (bar: LP-LD) to show performance impact with increase in page size.

\begin{figure}[b]
\begin{center}
\includegraphics[width=0.7\columnwidth]{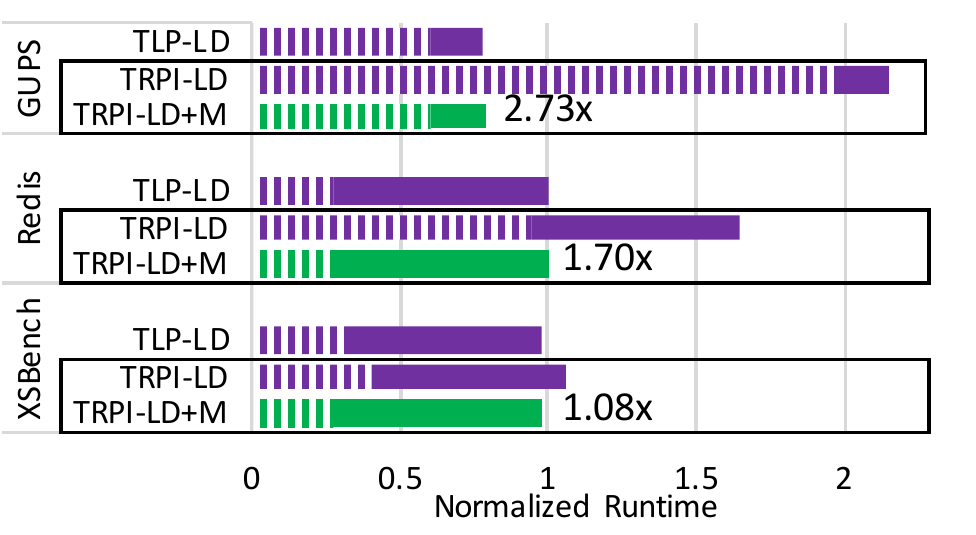}
\end{center}
        \vspace{-0.3cm}
        \caption{Performance of \sys in workload migration scenario with 2MB pages under heavy memory fragmentation.}
\label{fig:frag}
\end{figure}

With 4KB pages (\autoref{fig:eval:single:mitosis}), remote page-tables cause
1.4x to 3.2x slowdown (bar: RPI-LD) relative to the baseline (LP-LD). 
\sys can mitigate this overhead and has the same performance 
as the baseline by migrating the page-tables with process migration.

With 2MB large pages (\autoref{fig:eval:single2m:mitosis}), we see that the page
walk overheads are comparatively lower, nevertheless we
observe a slowdown of up to 2.3x for TRPI-LD over TLP-LD configuration. Again, \sys can
mitigate this overhead and has the same performance as the TLP-LD configuration.
Note, that for certain workloads the page-tables are cached well in the CPU caches 
and thus there is no difference in runtime. For example, in the case
of GUPS, we observe roughly one TLB miss per data access--two cache-line
requests in total per data array access. By breaking this down, we obtain that each
leaf page-table
cache-line covers about 16MB of memory which corresponds to 256k cache-lines of
the data array. Therefore, the page-table cache-lines are accessed 256k more
often than the data array cache-lines, and there are less than 500k page-table cache lines which
can easily be cached in L3 cache of the socket. In summary, page-table entries are
likely to be present in the sockets processor cache.

\noindent
{\bf Memory Fragmentation:} Physical memory fragmentation limits the availability
of large pages as the system ages, leading to higher page-walk overheads~\cite{ingens,ashish:2018}. \autoref{fig:frag} shows the performance of Mitosis under heavy
fragmentation while using THP in Linux with 2MB page size. We observe that all workloads, including those
that did not show performance improvement with \sys while using 2MB pages in \autoref{fig:eval:single2m:mitosis},
show dramatic improvement with \sys in this case. This is due to workloads falling back to 4KB pages under
fragmentation -- which we have already shown to be susceptible to NUMA effects of page-table walks.
Note that we present this experiment under heavy fragmentation to demonstrate that even if large pages are enabled, page-walk overheads can approach 
that of 4KB pages. In practice, the actual state of memory fragmentation may depend on several factors and these overheads will be proportional to the failure rate of large page allocations.

\noindent
{\bf Summary:} With this evaluation, we have shown that \sys completely avoids
resulting overheads due to page-tables being misplaced on
remote NUMA sockets. In none of the cases, \sys resulted in a slowdown of the
workload.

\subsection{Space and Runtime Overheads}
\label{sec:eval:memoryoverheads}
Enabling \sys implies maintaining replicas which consume memory and use CPU 
cycles to be kept consistent. We evaluate these overheads by estimating 
the additional memory requirement, and then perform micro-benchmarks on the 
virtual memory operations and wrap up by running applications end-to-end to set 
those overheads into perspective.

\subsubsection{Memory Overheads}

We estimate the overhead of the additional memory used to store the page-table 
replicas when \sys is enabled.
We define the two-dimensional function
\newline\hspace*{\fill}
$mem\_overhead(Footprint, Replicas) = Overhead\%$
\hspace*{\fill}\newline
that calculates memory overhead relative to the single page-table baseline and 
evaluate it using different values for the application's memory footprint 
and the number of replicas. For this estimation, we assume 4-level x86 paging 
with a compact address space e.g. the application uses addresses 
$0..FootPrint$. Each level has at least one page-table allocated and a page-table
 is 4KB in size.

\begin{table}[b]
  \begin{footnotesize}
   \begin{center}
  \begin{tabular}{cccccccc} \hline
    & & \multicolumn{5}{c}{{\bf Number of Replicas}} \\
    {\bf Footprint} & {\bf PT Size} & {\bf 1} & {\bf 2} & {\bf 4} & {\bf 8} & 
    {\bf 16} \\ \hline
    1 MB & 0.02 MB & 1.0 & 1.015 & 1.046 & 1.108 & 1.231 \\
    1 GB & 2.01 MB & 1.0 & 1.002 & 1.006 & 1.014 & 1.029 \\
    1 TB & 2.00 GB & 1.0 & 1.002 & 1.006 & 1.014 & 1.029 \\
   16 TB & 32.0 GB & 1.0 & 1.002 & 1.006 & 1.014 & 1.029 \\ \hline
  \end{tabular}
  \vspace{-0.3cm}
\end{center}
  \end{footnotesize}
  \caption{Memory footprint overhead for \sys}
  \label{tab:eval:memoverhead}
\end{table}

\autoref{tab:eval:memoverhead} shows the memory overheads of \sys for small to 
large applications using up to 16 replicas. We use the single page-table case as 
the baseline. The page-table accounts for about 0.19\% of the total footprint, 
except for the 1MB case where it accounts for 1.5\%. With an increasing memory 
footprint used by the application, \sys requires less than 2.9\% of additional 
memory for 16-replicas, whereas our four-socket machine used just 0.6\% 
additional memory.

The page-tables use a small fraction of the total memory footprint of the 
application. For small programs, the fraction is higher because there is a hard 
minimum of at least 16KB of page-tables--a 4KB page for each level. This is 
reflected by the large 23.1\% increase in memory consumption for small 
programs. However, putting this into perspective we advocate not to use \sys in 
this case as the 1MB memory footprint falls within the TLB coverage.

In summary, we showed that even with a 16-socket NUMA machine, \sys 
adds just 2.9\% memory overhead and this overhead drops to 0.6\% for our 
four-socket machine.

\subsubsection{VMA Operation Overheads}

In this part of the evaluation, we are interested in understanding the overheads of
self-replicating page-tables for common virtual memory operations such as \emph{mmap}, \emph{mprotect}
and \emph{munmap}.

%
We conducted a micro-benchmark that repeatedly calls the VMA operations and 
measured the time to complete the corresponding system calls. For each operation, we 
enforce that the page-table modifications are carried out e.g. by passing 
the~\texttt{MAP\_POPULATE} flat to \texttt{mmap}. We varied the number 
of affected pages from a single page to a large region of memory of multiple GB 
in size. We ran the micro-benchmark with \sys enabled and disabled on an 
otherwise idle system. We use 4KB pages and 4-way replication.

\begin{table}[t]
\begin{footnotesize}
\begin{center}
\begin{tabular}{cccc} \hline
    \textbf{Operation} & \textbf{4KB region} & \textbf{8MB region} & \textbf{4GB 
    region} \\ \hline
    mmap               &       1.021x &       1.008x &       1.006x \\
    mprotect           &       1.121x &       3.238x &       3.279x \\
    munmap             &       1.043x &       1.354x &       1.393x \\ \hline
\end{tabular}
\end{center}
\end{footnotesize}
\vspace{-0.3cm}
\caption{Runtime overhead of \sys for virtual memory operation system calls 
using 4-way Replication.}
\label{tab:eval:rtoverhead}
\end{table}

The results of this micro-benchmark are shown in~\autoref{tab:eval:rtoverhead}. 
The table shows CPU cycles required to perform the operation on a memory region of 
size 4KB, 8MB, or 4GB with \sys being on or off. Further, we calculate the 
overheads of \sys by dividing the 4-way replicated case (\sys on) with the base 
case, \sys off. For \texttt{mmap}, we observe an overhead of 
less than 2\%. For \texttt{unmap}, the overhead grows to 35\% while \sys 
adds more than 3x overheads for \texttt{mprotect}. 

\begin{table}[b]
  \begin{footnotesize}
    \begin{center}
    \begin{tabular}{cccc} \hline
      \textbf{Workload}& \textbf{\sys Off} &
      \textbf{\sys On} & \textbf{Overhead} \\ \hline
      GUPS  & 270.93 (0.43) &  272.18 (0.00) &  0.46\% \\
      Redis & 633.94 (0.34) &  636.31 (0.86) &  0.37\% \\ \hline
    \end{tabular}
    \end{center}
  \end{footnotesize}
  \vspace{-0.3cm}
  \caption{Runtimes with LP-LD setting, including initialization with and
  without \sys. Standard Deviation in Brackets.}
  \label{tab:eval:endtoend}
\end{table}

With 4-way replication, there are four sets of page-tables that need to be 
updated resulting in four times the work. We attribute the rather low overhead 
for \texttt{mmap} to the allocation and zeroing of new data pages during the 
system call. Likewise, when performing the \texttt{unmap} the freed pages are handed 
back to the allocator, but not zeroed resulting in less work per page and thus higher 
overhead of replication. \sys experiences a large overhead for 
\texttt{mprotect} which is still smaller than the replication factor. The 
\texttt{mprotect} operation does a read-modify-write cycle on the affected 
page-table entries. This process is efficient with no replicas as it results in 
sequential access within a page-table. However, with the PV-OPS interface, for 
each written entry all replicas are updated accordingly which kills locality. 
This can be avoided by either changing the PV-Ops interface or implementing lazy 
updates.

\subsubsection{No End-to-End Slowdown}
We now set the VMA operations micro-benchmark of the 
previous section into the perspective of real-world applications. We show that 
our modifications to the Linux kernel to support \sys has negligible end-to-end overhead 
for applications.

We compare the execution time of the single-threaded benchmarks. We 
run those benchmarks with and without \sys and measure overall execution time,
including  allocation and initialization phase. We use the LP-LD configuration, i.e. 
everything is locally allocated. THP is deactivated.

The results are shown in~\autoref{tab:eval:endtoend}. 
We observe that in both cases, GUPS and Redis, the overheads of \sys are less 
than half a percent, which is small compared to the improvements we have 
demonstrated earlier.

%% file: 80-conclusion.tex
\section{Conclusion}
\label{sec:conclusion}

We presented \sys: a technique that transparently replicates page-tables 
on large-memory machines, and provides the first platform to systematically evaluate
page-table allocation policies inside the OS. With strong empirical evidence, we
made the case for taking the allocation and placement of page-tables to a first-class consideration, 
in turn, optimizing performance on NUMA systems. We
also demonstrated the benefits of replicating  page-tables in large-memory
machines for various use-cases, while observing negligible memory and runtime
overheads. We plan to open-source the tools used in this work to inspire further
research on optimizing page-table placement. Moreover, we plan to work with the Linux community
to get \sys integrated into the mainline kernel.

%% file: paper-submission.bbl
\begin{thebibliography}{10}
\providecommand{\url}[1]{#1}
\csname url@samestyle\endcsname
\providecommand{\newblock}{\relax}
\providecommand{\bibinfo}[2]{#2}
\providecommand{\BIBentrySTDinterwordspacing}{\spaceskip=0pt\relax}
\providecommand{\BIBentryALTinterwordstretchfactor}{4}
\providecommand{\BIBentryALTinterwordspacing}{\spaceskip=\fontdimen2\font plus
\BIBentryALTinterwordstretchfactor\fontdimen3\font minus
  \fontdimen4\font\relax}
\providecommand{\BIBforeignlanguage}[2]{{%
\expandafter\ifx\csname l@#1\endcsname\relax
\typeout{** WARNING: IEEEtranS.bst: No hyphenation pattern has been}%
\typeout{** loaded for the language `#1'. Using the pattern for}%
\typeout{** the default language instead.}%
\else
\language=\csname l@#1\endcsname
\fi
#2}}
\providecommand{\BIBdecl}{\relax}
\BIBdecl

\bibitem{latency}
``Amd epyc infinity fabric latency ddr4 2400 v 2666: A snapshot,''
  \url{https://www.servethehome.com/amd-epyc-infinity-fabric-latency-ddr4-2400-v-2666-a-snapshot/}.

\bibitem{autonuma}
``{AutoNUMA: the other approach to NUMA scheduling},''
  \url{https://lwn.net/articles/488709/}.

\bibitem{vmware:numa}
``{Extreme Performance Series: vSphere Compute \& Memory Schedulers},''
  \url{https://static.rainfocus.com/vmware/vmworldus17/sess/1489512432328001AfWH/finalpresentationPDF/SER2343BU_FORMATTED_FINAL_1507912874739001gpDS.pdf}.

\bibitem{intel:eptadbits}
``{FOUR NEW VIRTUALIZATION TECHNOLOGIES ON THE LATEST INTEL® XEON},''
  \url{https://software.intel.com/en-us/blogs/2014/09/08/four-new-virtualization-technologies-on-the-latest-intel-xeon-are-you-ready-to}.

\bibitem{graph500}
``{Graph500 | large scale benchmarks},'' \url{https://graph500.org}.

\bibitem{intel:3dxpoint}
``{Intel’s Enterprise Extravaganza 2019: Launching Cascade Lake, Optane
  DCPMM, Agilex FPGAs, 100G Ethernet, and Xeon D-1600},''
  \url{https://www.anandtech.com/show/14155/intels-enterprise-extravaganza-2019-roundup}.

\bibitem{svm}
``Liblinear -- a library for large linear classification,''
  \url{https://www.csie.ntu.edu.tw/~cjlin/liblinear/}.

\bibitem{memcached}
``{memcached: a distributed memory object caching system},''
  \url{https://memcached.org}.

\bibitem{pvops}
``{Paravirt\_ops},''
  \url{https://www.kernel.org/doc/Documentation/virtual/paravirt\_ops.txt}.

\bibitem{parsec}
``Parsec benchmark suite,'' \url{https://parsec.cs.princeton.edu/overview.htm}.

\bibitem{gups}
``{RandomAccess: GUPS (Giga Updates Per Second)},''
  \url{https://icl.utk.edu/projectsfiles/hpcc/RandomAccess/}.

\bibitem{redis}
``{Redis},'' \url{https://redis.io}.

\bibitem{stream}
``{STREAM: Sustainable Memory Bandwidth in High Performance Computers},''
  \url{https://www.cs.virginia.edu/stream/}.

\bibitem{xsbench}
``{XSBench: The Monte Carlo Macroscopic Cross Section Lookup Benchmark},''
  \url{https://github.com/ANL-CESAR/XSBench}.

\bibitem{xv6}
``{Xv6, a simple Unix-like teaching operating system},''
  \url{https://pdos.csail.mit.edu/6.828/2012/xv6.html}.

\bibitem{Alam:2017}
H.~Alam, T.~Zhang, M.~Erez, and Y.~Etsion, ``Do-it-yourself virtual memory
  translation,'' in \emph{Proceedings of the 44th Annual International
  Symposium on Computer Architecture}, ser. ISCA '17, 2017, pp. 457--468.

\bibitem{amd:nextgen}
{AMD}, ``{The Next Generation AMD Enterprise Server Product Architecture},''
  \url{https://www.hotchips.org/wp-content/uploads/hc_archives/hc29/HC29.22-Tuesday-Pub/HC29.22.90-Server-Pub/HC29.22.921-EPYC-Lepak-AMD-v2.pdf}.

\bibitem{Barham:2003:XAV}
\BIBentryALTinterwordspacing
P.~Barham, B.~Dragovic, K.~Fraser, S.~Hand, T.~Harris, A.~Ho, R.~Neugebauer,
  I.~Pratt, and A.~Warfield, ``{Xen and the Art of Virtualization},'' in
  \emph{Proceedings of the Nineteenth ACM Symposium on Operating Systems
  Principles}, ser. SOSP '03.\hskip 1em plus 0.5em minus 0.4em\relax Bolton
  Landing, NY, USA: ACM, 2003, pp. 164--177. [Online]. Available:
  \url{http://doi.acm.org/10.1145/945445.945462}
\BIBentrySTDinterwordspacing

\bibitem{Barr:2010}
\BIBentryALTinterwordspacing
T.~W. Barr, A.~L. Cox, and S.~Rixner, ``{Translation Caching: Skip, Don't Walk
  (the Page Table)},'' in \emph{Proceedings of the 37th Annual International
  Symposium on Computer Architecture}, ser. ISCA '10, Saint-Malo, France, 2010,
  pp. 48--59. [Online]. Available:
  \url{http://doi.acm.org/10.1145/1815961.1815970}
\BIBentrySTDinterwordspacing

\bibitem{Barr:2011}
T.~W. Barr, A.~L. Cox, and S.~Rixner, ``{SpecTLB: A mechanism for speculative
  address translation},'' in \emph{2011 38th Annual International Symposium on
  Computer Architecture (ISCA)}, June 2011, pp. 307--317.

\bibitem{Basu:2013:dir_seg}
\BIBentryALTinterwordspacing
A.~Basu, J.~Gandhi, J.~Chang, M.~D. Hill, and M.~M. Swift, ``{Efficient Virtual
  Memory for Big Memory Servers},'' in \emph{Proceedings of the 40th Annual
  International Symposium on Computer Architecture}, ser. ISCA '13, Tel-Aviv,
  Israel, 2013, pp. 237--248. [Online]. Available:
  \url{http://doi.acm.org/10.1145/2485922.2485943}
\BIBentrySTDinterwordspacing

\bibitem{Baumann:2009}
\BIBentryALTinterwordspacing
A.~Baumann, P.~Barham, P.-E. Dagand, T.~Harris, R.~Isaacs, S.~Peter, T.~Roscoe,
  A.~Sch\"{u}pbach, and A.~Singhania, ``{The Multikernel: A New OS Architecture
  for Scalable Multicore Systems},'' in \emph{Proceedings of the ACM SIGOPS
  22Nd Symposium on Operating Systems Principles}, ser. SOSP '09, Big Sky,
  Montana, USA, 2009, pp. 29--44. [Online]. Available:
  \url{http://doi.acm.org/10.1145/1629575.1629579}
\BIBentrySTDinterwordspacing

\bibitem{gapbs}
\BIBentryALTinterwordspacing
S.~Beamer, K.~Asanovic, and D.~A. Patterson, ``The {GAP} benchmark suite,''
  \emph{CoRR}, vol. abs/1508.03619, 2015. [Online]. Available:
  \url{http://arxiv.org/abs/1508.03619}
\BIBentrySTDinterwordspacing

\bibitem{Bhattacharjee:2013}
\BIBentryALTinterwordspacing
A.~Bhattacharjee, ``{Large-reach Memory Management Unit Caches},'' in
  \emph{Proceedings of the 46th Annual IEEE/ACM International Symposium on
  Microarchitecture}, ser. MICRO-46, Davis, California, 2013, pp. 383--394.
  [Online]. Available: \url{http://doi.acm.org/10.1145/2540708.2540741}
\BIBentrySTDinterwordspacing

\bibitem{Bhattacharjee:2017}
\BIBentryALTinterwordspacing
A.~Bhattacharjee, ``{Translation-Triggered Prefetching},'' in \emph{Proceedings
  of the Twenty-Second International Conference on Architectural Support for
  Programming Languages and Operating Systems}, ser. ASPLOS '17, Xi'an, China,
  2017, pp. 63--76. [Online]. Available:
  \url{http://doi.acm.org/10.1145/3037697.3037705}
\BIBentrySTDinterwordspacing

\bibitem{Bhattacharjee:2011}
\BIBentryALTinterwordspacing
A.~Bhattacharjee, D.~Lustig, and M.~Martonosi, ``{Shared Last-level TLBs for
  Chip Multiprocessors},'' in \emph{Proceedings of the 2011 IEEE 17th
  International Symposium on High Performance Computer Architecture}, ser. HPCA
  '11, 2011, pp. 62--63. [Online]. Available:
  \url{http://dl.acm.org/citation.cfm?id=2014698.2014896}
\BIBentrySTDinterwordspacing

\bibitem{Bouron:2018:BSF}
\BIBentryALTinterwordspacing
J.~Bouron, S.~Chevalley, B.~Lepers, W.~Zwaenepoel, R.~Gouicem, J.~Lawall,
  G.~Muller, and J.~Sopena, ``{The Battle of the Schedulers: FreeBSD ULE vs.
  Linux CFS},'' in \emph{Proceedings of the 2018 USENIX Conference on Usenix
  Annual Technical Conference}, ser. USENIX ATC '18.\hskip 1em plus 0.5em minus
  0.4em\relax Berkeley, CA, USA: USENIX Association, 2018, pp. 85--96.
  [Online]. Available: \url{http://dl.acm.org/citation.cfm?id=3277355.3277364}
\BIBentrySTDinterwordspacing

\bibitem{Boyd-Wickizer:2008}
\BIBentryALTinterwordspacing
S.~Boyd-Wickizer, H.~Chen, R.~Chen, Y.~Mao, F.~Kaashoek, R.~Morris,
  A.~Pesterev, L.~Stein, M.~Wu, Y.~Dai, Y.~Zhang, and Z.~Zhang, ``{Corey: An
  Operating System for Many Cores},'' in \emph{Proceedings of the 8th USENIX
  Conference on Operating Systems Design and Implementation}, ser.
  OSDI'08.\hskip 1em plus 0.5em minus 0.4em\relax San Diego, California: USENIX
  Association, 2008, pp. 43--57. [Online]. Available:
  \url{http://dl.acm.org/citation.cfm?id=1855741.1855745}
\BIBentrySTDinterwordspacing

\bibitem{Calciu:2017:BCD}
\BIBentryALTinterwordspacing
I.~Calciu, S.~Sen, M.~Balakrishnan, and M.~K. Aguilera, ``{Black-box Concurrent
  Data Structures for NUMA Architectures},'' in \emph{Proceedings of the
  Twenty-Second International Conference on Architectural Support for
  Programming Languages and Operating Systems}, ser. ASPLOS '17, Xi'an, China,
  2017, pp. 207--221. [Online]. Available:
  \url{http://doi.acm.org/10.1145/3037697.3037721}
\BIBentrySTDinterwordspacing

\bibitem{Clements:2013}
A.~T. Clements, M.~F. Kaashoek, and N.~Zeldovich, ``{RadixVM: Scalable Address
  Spaces for Multithreaded Applications},'' in \emph{Proceedings of the 8th ACM
  European Conference on Computer Systems}, ser. EuroSys '13, Prague, Czech
  Republic, 2013, pp. 211--224.

\bibitem{Das:Albatross}
S.~Das, S.~Nishimura, D.~Agrawal, and A.~El~Abbadi, ``Albatross: Lightweight
  elasticity in shared storage databases for the cloud using live data
  migration,'' in \emph{Proceedings of the 2011 VLDB Endowment}, ser. VLDB '11,
  2011.

\bibitem{Dashti:2013}
\BIBentryALTinterwordspacing
M.~Dashti, A.~Fedorova, J.~Funston, F.~Gaud, R.~Lachaize, B.~Lepers, V.~Quema,
  and M.~Roth, ``{Traffic Management: A Holistic Approach to Memory Placement
  on NUMA Systems},'' in \emph{Proceedings of the Eighteenth International
  Conference on Architectural Support for Programming Languages and Operating
  Systems}, ser. ASPLOS '13, Houston, Texas, USA, 2013, pp. 381--394. [Online].
  Available: \url{http://doi.acm.org/10.1145/2451116.2451157}
\BIBentrySTDinterwordspacing

\bibitem{Demir:2014}
\BIBentryALTinterwordspacing
Y.~Demir, Y.~Pan, S.~Song, N.~Hardavellas, J.~Kim, and G.~Memik, ``{Galaxy: A
  High-performance Energy-efficient Multi-chip Architecture Using Photonic
  Interconnects},'' in \emph{Proceedings of the 28th ACM International
  Conference on Supercomputing}, ser. ICS '14, Munich, Germany, 2014, pp.
  303--312. [Online]. Available:
  \url{http://doi.acm.org/10.1145/2597652.2597664}
\BIBentrySTDinterwordspacing

\bibitem{Du:2015}
Y.~Du, M.~Zhou, B.~R. Childers, D.~Mossé, and R.~Melhem, ``{Supporting
  Superpages in Non-Contiguous Physical Memory},'' in \emph{2015 IEEE 21st
  International Symposium on High Performance Computer Architecture (HPCA)},
  Feb 2015, pp. 223--234.

\bibitem{Fang:2001}
\BIBentryALTinterwordspacing
Z.~Fang, L.~Zhang, J.~B. Carter, W.~C. Hsieh, and S.~A. McKee, ``{Reevaluating
  Online Superpage Promotion with Hardware Support},'' in \emph{Proceedings of
  the 7th International Symposium on High-Performance Computer Architecture},
  ser. HPCA '01, 2001, pp. 63--. [Online]. Available:
  \url{http://dl.acm.org/citation.cfm?id=580550.876428}
\BIBentrySTDinterwordspacing

\bibitem{Ganapathy:1998}
\BIBentryALTinterwordspacing
N.~Ganapathy and C.~Schimmel, ``{General Purpose Operating System Support for
  Multiple Page Sizes},'' in \emph{Proceedings of the Annual Conference on
  USENIX Annual Technical Conference}, ser. ATEC '98, New Orleans, Louisiana,
  1998, pp. 8--8. [Online]. Available:
  \url{http://dl.acm.org/citation.cfm?id=1268256.1268264}
\BIBentrySTDinterwordspacing

\bibitem{Gandhi:2017:micro}
J.~Gandhi, M.~D. Hill, and M.~M. Swift, ``{Agile Paging for Efficient Memory
  Virtualization},'' \emph{IEEE Micro}, vol.~37, no.~3, pp. 80--86, 2017.

\bibitem{Gandhi:2016:micro}
J.~Gandhi, V.~Karakostas, F.~Ayar, A.~Cristal, M.~D. Hill, K.~S. McKinley,
  M.~Nemirovsky, M.~M. Swift, and O.~S. Ünsal, ``{Range Translations for Fast
  Virtual Memory},'' \emph{IEEE Micro}, vol.~36, no.~3, pp. 118--126, May 2016.

\bibitem{Gandhi:2014:emv}
\BIBentryALTinterwordspacing
J.~Gandhi, A.~Basu, M.~D. Hill, and M.~M. Swift, ``{Efficient Memory
  Virtualization: Reducing Dimensionality of Nested Page Walks},'' in
  \emph{Proceedings of the 47th Annual IEEE/ACM International Symposium on
  Microarchitecture}, ser. MICRO-47, Cambridge, United Kingdom, 2014, pp.
  178--189. [Online]. Available: \url{http://dx.doi.org/10.1109/MICRO.2014.37}
\BIBentrySTDinterwordspacing

\bibitem{Gandhi:2016:agile}
\BIBentryALTinterwordspacing
J.~Gandhi, M.~D. Hill, and M.~M. Swift, ``{Agile Paging: Exceeding the Best of
  Nested and Shadow Paging},'' in \emph{Proceedings of the 43rd International
  Symposium on Computer Architecture}, ser. ISCA '16, Seoul, Republic of Korea,
  2016, pp. 707--718. [Online]. Available:
  \url{https://doi.org/10.1109/ISCA.2016.67}
\BIBentrySTDinterwordspacing

\bibitem{Gaud:2014:LPM}
\BIBentryALTinterwordspacing
F.~Gaud, B.~Lepers, J.~Decouchant, J.~Funston, A.~Fedorova, and V.~Qu{\'e}ma,
  ``{Large Pages May Be Harmful on NUMA Systems},'' in \emph{Proceedings of the
  2014 USENIX Conference on USENIX Annual Technical Conference}, ser. USENIX
  ATC'14, Philadelphia, PA, 2014, pp. 231--242. [Online]. Available:
  \url{http://dl.acm.org/citation.cfm?id=2643634.2643659}
\BIBentrySTDinterwordspacing

\bibitem{Haria:2018}
\BIBentryALTinterwordspacing
S.~Haria, M.~D. Hill, and M.~M. Swift, ``Devirtualizing memory in heterogeneous
  systems,'' in \emph{Proceedings of the Twenty-Third International Conference
  on Architectural Support for Programming Languages and Operating Systems},
  ser. ASPLOS '18.\hskip 1em plus 0.5em minus 0.4em\relax New York, NY, USA:
  ACM, 2018, pp. 637--650. [Online]. Available:
  \url{http://doi.acm.org/10.1145/3173162.3173194}
\BIBentrySTDinterwordspacing

\bibitem{intel:5level}
{Intel Corp.}, ``{5-Level Paging and 5-Level EPT},''
  \url{https://software.intel.com/sites/default/files/managed/2b/80/5-level_paging_white_paper.pdf}.

\bibitem{intel:amd}
{Intel Corp.}, ``{New Intel Core Processor Combines High-Performance CPU with
  Custom Discrete Graphics from AMD to Enable Sleeker, Thinner Devices},''
  \url{https://newsroom.intel.com/editorials/new-intel-core-processor-combine-high-performance-cpu-discrete-graphics-sleek-thin-devices/}.

\bibitem{Iyer:2016}
S.~S. Iyer, ``{Heterogeneous Integration for Performance and Scaling},''
  \emph{IEEE Transactions on Components, Packaging and Manufacturing
  Technology}, vol.~6, no.~7, pp. 973--982, July 2016.

\bibitem{Kaestle:2015:SSA}
\BIBentryALTinterwordspacing
S.~Kaestle, R.~Achermann, T.~Roscoe, and T.~Harris, ``{Shoal: Smart Allocation
  and Replication of Memory for Parallel Programs},'' in \emph{Proceedings of
  the 2015 USENIX Conference on Usenix Annual Technical Conference}, ser.
  USENIX ATC '15, Santa Clara, CA, 2015, pp. 263--276. [Online]. Available:
  \url{http://dl.acm.org/citation.cfm?id=2813767.2813787}
\BIBentrySTDinterwordspacing

\bibitem{Kandiraju:2002}
\BIBentryALTinterwordspacing
G.~B. Kandiraju and A.~Sivasubramaniam, ``{Going the Distance for TLB
  Prefetching: An Application-driven Study},'' in \emph{Proceedings of the 29th
  Annual International Symposium on Computer Architecture}, ser. ISCA '02,
  2002, pp. 195--206. [Online]. Available:
  \url{http://dl.acm.org/citation.cfm?id=545215.545237}
\BIBentrySTDinterwordspacing

\bibitem{Kannan:2015}
A.~Kannan, N.~E. Jerger, and G.~H. Loh, ``{Enabling interposer-based
  disintegration of multi-core processors},'' in \emph{2015 48th Annual
  IEEE/ACM International Symposium on Microarchitecture (MICRO)}, Dec 2015, pp.
  546--558.

\bibitem{Karakostas:2014}
V.~Karakostas, O.~S. Unsal, M.~Nemirovsky, A.~Cristal, and M.~Swift,
  ``{Performance analysis of the memory management unit under scale-out
  workloads},'' in \emph{2014 IEEE International Symposium on Workload
  Characterization (IISWC)}, Oct 2014, pp. 1--12.

\bibitem{Karakostas:2015:rmm}
\BIBentryALTinterwordspacing
V.~Karakostas, J.~Gandhi, F.~Ayar, A.~Cristal, M.~D. Hill, K.~S. McKinley,
  M.~Nemirovsky, M.~M. Swift, and O.~\"{U}nsal, ``{Redundant Memory Mappings
  for Fast Access to Large Memories},'' in \emph{Proceedings of the 42Nd Annual
  International Symposium on Computer Architecture}, ser. ISCA '15, Portland,
  Oregon, 2015, pp. 66--78. [Online]. Available:
  \url{http://doi.acm.org/10.1145/2749469.2749471}
\BIBentrySTDinterwordspacing

\bibitem{ingens}
\BIBentryALTinterwordspacing
Y.~Kwon, H.~Yu, S.~Peter, C.~J. Rossbach, and E.~Witchel, ``Coordinated and
  efficient huge page management with ingens,'' in \emph{Proceedings of the
  12th USENIX Conference on Operating Systems Design and Implementation}, ser.
  OSDI'16.\hskip 1em plus 0.5em minus 0.4em\relax Berkeley, CA, USA: USENIX
  Association, 2016, pp. 705--721. [Online]. Available:
  \url{http://dl.acm.org/citation.cfm?id=3026877.3026931}
\BIBentrySTDinterwordspacing

\bibitem{Lozi:2016:LSD}
\BIBentryALTinterwordspacing
J.-P. Lozi, B.~Lepers, J.~Funston, F.~Gaud, V.~Qu{\'e}ma, and A.~Fedorova,
  ``The linux scheduler: A decade of wasted cores,'' in \emph{Proceedings of
  the Eleventh European Conference on Computer Systems}, ser. EuroSys
  '16.\hskip 1em plus 0.5em minus 0.4em\relax New York, NY, USA: ACM, 2016, pp.
  1:1--1:16. [Online]. Available:
  \url{http://doi.acm.org/10.1145/2901318.2901326}
\BIBentrySTDinterwordspacing

\bibitem{Lustig:2013}
\BIBentryALTinterwordspacing
D.~Lustig, A.~Bhattacharjee, and M.~Martonosi, ``{TLB Improvements for Chip
  Multiprocessors: Inter-Core Cooperative Prefetchers and Shared Last-Level
  TLBs},'' \emph{ACM Trans. Archit. Code Optim.}, vol.~10, no.~1, pp.
  2:1--2:38, Apr. 2013. [Online]. Available:
  \url{http://doi.acm.org/10.1145/2445572.2445574}
\BIBentrySTDinterwordspacing

\bibitem{mochi}
{Marvell Corporation}, ``{MoChi Architecture},''
  \url{http://www.marvell.com/architecture/mochi/}.

\bibitem{Navarro:2002}
\BIBentryALTinterwordspacing
J.~Navarro, S.~Iyer, P.~Druschel, and A.~Cox, ``{Practical, Transparent
  Operating System Support for Superpages},'' \emph{SIGOPS Oper. Syst. Rev.},
  vol.~36, no.~SI, pp. 89--104, Dec. 2002. [Online]. Available:
  \url{http://doi.acm.org/10.1145/844128.844138}
\BIBentrySTDinterwordspacing

\bibitem{ashish:2018}
\BIBentryALTinterwordspacing
A.~Panwar, A.~Prasad, and K.~Gopinath, ``Making huge pages actually useful,''
  in \emph{Proceedings of the Twenty-Third International Conference on
  Architectural Support for Programming Languages and Operating Systems}, ser.
  ASPLOS '18.\hskip 1em plus 0.5em minus 0.4em\relax New York, NY, USA: ACM,
  2018, pp. 679--692. [Online]. Available:
  \url{http://doi.acm.org/10.1145/3173162.3173203}
\BIBentrySTDinterwordspacing

\bibitem{Papadopoulou:2015}
M.~Papadopoulou, X.~Tong, A.~Seznec, and A.~Moshovos, ``{Prediction-based
  superpage-friendly TLB designs},'' in \emph{2015 IEEE 21st International
  Symposium on High Performance Computer Architecture (HPCA)}, Feb 2015, pp.
  210--222.

\bibitem{Pham:2014}
B.~Pham, A.~Bhattacharjee, Y.~Eckert, and G.~H. Loh, ``{Increasing TLB reach by
  exploiting clustering in page translations},'' in \emph{2014 IEEE 20th
  International Symposium on High Performance Computer Architecture (HPCA)},
  Feb 2014, pp. 558--567.

\bibitem{Pham:2012}
\BIBentryALTinterwordspacing
B.~Pham, V.~Vaidyanathan, A.~Jaleel, and A.~Bhattacharjee, ``{CoLT: Coalesced
  Large-Reach TLBs},'' in \emph{Proceedings of the 2012 45th Annual IEEE/ACM
  International Symposium on Microarchitecture}, ser. MICRO-45, Vancouver,
  B.C., CANADA, 2012, pp. 258--269. [Online]. Available:
  \url{https://doi.org/10.1109/MICRO.2012.32}
\BIBentrySTDinterwordspacing

\bibitem{Pham:2015}
\BIBentryALTinterwordspacing
B.~Pham, J.~Vesel\'{y}, G.~H. Loh, and A.~Bhattacharjee, ``{Large Pages and
  Lightweight Memory Management in Virtualized Environments: Can You Have It
  Both Ways?}'' in \emph{Proceedings of the 48th International Symposium on
  Microarchitecture}, ser. MICRO-48, Waikiki, Hawaii, 2015, pp. 1--12.
  [Online]. Available: \url{http://doi.acm.org/10.1145/2830772.2830773}
\BIBentrySTDinterwordspacing

\bibitem{Rangan:Thread}
K.~Rangan, G.-Y. Wei, and D.~Brooks, ``Thread motion: Fine-grained power
  management for multi-core systems,'' in \emph{Proceedings of the 2009
  International Symposium on Computer Architecture}, ser. ISCA '09, 2009.

\bibitem{Saulsbury:2000}
\BIBentryALTinterwordspacing
A.~Saulsbury, F.~Dahlgren, and P.~Stenstr\"{o}m, ``{Recency-based TLB
  Preloading},'' in \emph{Proceedings of the 27th Annual International
  Symposium on Computer Architecture}, ser. ISCA '00, Vancouver, British
  Columbia, Canada, 2000, pp. 117--127. [Online]. Available:
  \url{http://doi.acm.org/10.1145/339647.339666}
\BIBentrySTDinterwordspacing

\bibitem{Seznec:2004}
\BIBentryALTinterwordspacing
A.~Seznec, ``{Concurrent Support of Multiple Page Sizes on a Skewed Associative
  TLB},'' \emph{IEEE Trans. Comput.}, vol.~53, no.~7, pp. 924--927, Jul. 2004.
  [Online]. Available: \url{https://doi.org/10.1109/TC.2004.21}
\BIBentrySTDinterwordspacing

\bibitem{Swanson:1998}
\BIBentryALTinterwordspacing
M.~Swanson, L.~Stoller, and J.~Carter, ``{Increasing TLB Reach Using Superpages
  Backed by Shadow Memory},'' in \emph{Proceedings of the 25th Annual
  International Symposium on Computer Architecture}, ser. ISCA '98, Barcelona,
  Spain, 1998, pp. 204--213. [Online]. Available:
  \url{https://doi.org/10.1145/279358.279388}
\BIBentrySTDinterwordspacing

\bibitem{tsmc:packaging}
{Taiwan Semiconductor Manufacturing Company}, ``{CoWoS Services},''
  \url{http://www.tsmc.com/english/dedicatedFoundry/services/cowos.htm}.

\bibitem{Talluri:1994}
\BIBentryALTinterwordspacing
M.~Talluri and M.~D. Hill, ``{Surpassing the TLB Performance of Superpages with
  Less Operating System Support},'' in \emph{Proceedings of the Sixth
  International Conference on Architectural Support for Programming Languages
  and Operating Systems}, ser. ASPLOS VI, San Jose, California, USA, 1994, pp.
  171--182. [Online]. Available: \url{http://doi.acm.org/10.1145/195473.195531}
\BIBentrySTDinterwordspacing

\bibitem{Yin:2018:chiplet}
J.~Yin, Z.~Lin, O.~Kayiran, M.~Poremba, M.~S.~B. Altaf, N.~E. Jerger, and G.~H.
  Loh, ``{Modular Routing Design for Chiplet-Based Systems},'' in \emph{2018
  ACM/IEEE 45th Annual International Symposium on Computer Architecture
  (ISCA)}, June 2018, pp. 726--738.

\end{thebibliography}
